\pdfoutput=1
\RequirePackage{ifpdf}
\ifpdf % We are running pdfTeX in pdf mode
\documentclass[pdftex]{sigma}
\else
\documentclass{sigma}
\fi

\numberwithin{equation}{section}

\newtheorem{Theorem}{Theorem}[section]
\newtheorem{Proposition}[Theorem]{Proposition}
 { \theoremstyle{definition}
\newtheorem{Remark}[Theorem]{Remark} }

\begin{document}
\allowdisplaybreaks

\newcommand{\arXivNumber}{1902.08111}

\renewcommand{\PaperNumber}{079}

\FirstPageHeading

\ShortArticleName{Dispersionless Multi-Dimensional Integrable Systems}

\ArticleName{Dispersionless Multi-Dimensional Integrable Systems\\ and Related
Conformal Structure Generating\\ Equations of Mathematical Physics}

\Author{Oksana Ye.~HENTOSH~$^{\dag^1}$, Yarema A.~PRYKARPATSKY~$^{\dag^2\dag^3}$, Denis BLACKMORE~$^{\dag^4}$\\ and Anatolij K.~PRYKARPATSKI~$^{\dag^5}$}

\AuthorNameForHeading{O.Ye.~Hentosh, Ya.A.~Prykarpatsky, D.~Blackmore and A.K.~Prykarpatski}

\Address{$^{\dag^1}$~Pidstryhach Institute for Applied Problems of Mechanics and Mathematics of NAS of Ukraine,\\
\hphantom{$^{\dag^1}$}~Lviv, 79060, Ukraine}
\EmailDD{\href{mailto:ohen@ukr.net}{ohen@ukr.net}}

\Address{$^{\dag^2}$~Department of Applied Mathematics, University of Agriculture in Krakow, 30059, Poland}
\EmailDD{\href{mailto:yarpry@gmail.com}{yarpry@gmail.com}}

\Address{$^{\dag^3}$~Institute of Mathematics of NAS of Ukraine, Kyiv, 01024, Ukraine}

\Address{$^{\dag^4}$~Department of Mathematical Sciences, New Jersey Institute of Technology, \\
\hphantom{$^{\dag^4}$}~University Heights, Newark, NJ 07102 USA}
\EmailDD{\href{mailto:denis.l.blackmore@njit.edu}{denis.l.blackmore@njit.edu}}

\Address{$^{\dag^5}$~Department of Physics, Mathematics and Computer Science,\\
\hphantom{$^{\dag^5}$}~Cracow University of Technology, Cracow, 31155, Poland}
\EmailDD{\href{mailto:pryk.anat@cybergal.com}{pryk.anat@cybergal.com}}

\ArticleDates{Received April 08, 2019, in final form October 07, 2019; Published online October 14, 2019}

\Abstract{Using diffeomorphism group vector fields on $\mathbb{C}$-multiplied tori and the related Lie-algebraic structures, we study multi-dimensional dispersionless integrable systems that describe conformal structure generating equations of mathematical physics. An interesting modification of the devised Lie-algebraic approach subject to spatial-dimensional invariance and meromorphicity of the related differential-geometric structures is described and applied in proving complete integrability of some conformal structure generating equations. As examples, we analyze the Einstein--Weyl metric equation, the modified Einstein--Weyl metric equation, the Dunajski heavenly equation system, the first and second conformal structure generating equations and the inverse first Shabat reduction heavenly equation. We also analyze the modified Pleba\'nski heavenly equations, the Husain heavenly equation and the general Monge equation along with their multi-dimensional generalizations. In addition, we construct superconformal analogs of the Whitham heavenly equation.}


\Classification{17B68; 17B80; 35Q53; 35G25; 35N10; 37K35; 58J70; 58J72; 34A34; 37K05; 37K10}

\section[Vector fields on $\mathbb{T}^{n}\times\mathbb{C}$ and their related Lie-algebraic properties]{Vector fields on $\boldsymbol{\mathbb{T}^{n}\times\mathbb{C}}$\\ and their related Lie-algebraic properties} \label{Sec_1}

Consider the loop group $\tilde{G}:=\widetilde{{\rm Diff}}\big(\mathbb{T}^{n} \times\mathbb{C}\big)$, where $\mathbb{T}^{n}$ is the $n$-torus and $\mathbb{C}$ the complex plane, which is of the type in \cite{AdHaPr,Harn-3,Harn-1,Harn-2,PrSe}. In particular, it is the set of smooth mappings $\big\{\mathbb{C}^{1}\supset\mathbb{S}^{1}\longrightarrow
G := {\rm Diff}\big(\mathbb{T}^{n}\big)\big\}$, extended holomorphically to the interior $\mathbb{D}_{+}^{1}$ and exterior $\mathbb{D}_{-}^{1}=\mathbb{C} \backslash\overline{\mathbb{D}}_{+}^{1}$ of the unit circle $\mathbb{S}^{1}\subset\mathbb{C}$ in such a way that for any $g(\lambda)\in\tilde{G}$ the reality condition $\overline{g(\lambda)} =g(\overline{\lambda})$ holds for any $\lambda\in\mathbb{C}$. The corresponding diffeomorphism Lie algebra splitting $\tilde{\mathcal{G}}:= \tilde{\mathcal{G}}_{+}\oplus\mathcal{\tilde {G}}_{-}$, where $\tilde{\mathcal{G}}_{+}:= \widetilde{{\rm diff}} \big(\mathbb{T}^{n}\times\mathbb{C}\big)_{+}\subset \Gamma\big(\mathbb{T}^{n} \times\mathbb{C};T\big(\mathbb{T}^{n}\times\mathbb{C}\big)\big)$, is a Lie subalgebra comprising vector fields on the set $\mathbb{T}^{n}\times\mathbb{C}$, holomorphic on the disc $\mathbb{D}_{+}^{1}$ and $\mathcal{\tilde{G}}_{-}:= \widetilde{{\rm diff}}\big(\mathbb{T}^{n}\times\mathbb{C}\big)_{-} \subset  \Gamma\big(\mathbb{T}^{n}\times \mathbb{C};T\big(\mathbb{T}^{n}\times\mathbb{C}\big)\big)$ is a Lie subalgebra, consisting of vector fields on $\mathbb{T}^{n}\times\mathbb{C}$, holomorphic on $\mathbb{D}_{-}^{1}$. The adjoint space $\tilde{\mathcal{G}}^{\ast}:=\tilde{\mathcal{G}}_{+}^{\ast} \oplus\tilde{\mathcal{G}}_{-}^{\ast}$, where $\tilde{\mathcal{G}}_{+}^{\ast}\subset\Gamma\big(\mathbb{T}^{n} \times\mathbb{C};T^{\ast}\big(\mathbb{T}^{n}\times\mathbb{C}\big)\big)$ consists of the differential forms on the $\mathbb{T}^{n}\times\mathbb{C}$, holomorphic on the set $\mathbb{C}\backslash \overline{\mathbb{D}}_{+}^{1}$. Similarly, the adjoint space $\tilde{\mathcal{G}}_{-}^{\ast}\subset\Gamma\big(\mathbb{T}^{n}\times \mathbb{C};T^{\ast}\big(\mathbb{T}^{n}\times\mathbb{C}\big)\big)$ comprises the differential forms on the set $\mathbb{T}^{n}\times\mathbb{C}$ holomorphic on $\mathbb{D}_{+}^{1}$, so that the space $\tilde{\mathcal{G}}_{+}^{\ast}$ is dual to $\tilde{\mathcal{G}}_{+}$ and $\tilde{\mathcal{G}}_{-}^{\ast}$ is dual to $\tilde{\mathcal{G}}_{-}$ with respect to the convolution form on the product $\tilde{\mathcal{G}}^{\ast}\times\tilde{\mathcal{G}}$,
\begin{gather}
\big(\tilde{l}|\tilde{a}\big):=\operatorname{res}_{\lambda}\int_{\mathbb{T}^{n}}\langle l,a\rangle {\rm d}x,\label{eq1.1}
\end{gather}
for any vector field $\tilde{a}:=\big\langle a(\mathrm{x}),\frac{\partial}{\partial \mathrm{x}}\big\rangle \in\tilde{\mathcal{G}}$ and differential form $\tilde{l}:=\langle l(\mathrm{x}),{\rm d}\mathrm{x}\rangle \in\tilde{\mathcal{G}}^{\ast}$ on $\mathbb{T}^{n}\times\mathbb{C}$, depending on the coordinate $\mathrm{x} :=(\lambda;x)\in \mathbb{C} \times \mathbb{T}^{n}$. Here $\langle \cdot,\cdot\rangle $ is the usual scalar product on the Euclidean space $\mathbb{E}^{n+1}$ and $\frac{\partial}{\partial \mathrm{x}}:= \big(\frac{\partial}{\partial\lambda},\frac{\partial}{\partial x_{1}},\frac{\partial}{\partial x_{2}},\dots,\frac{\partial}{\partial x_{n}}\big)^{\top}$ is the gradient vector. The Lie algebra~$\tilde{\mathcal{G}}$ allows the direct sum splitting $\tilde{\mathcal{G}}=\tilde{\mathcal{G}}_{+}\oplus\tilde{\mathcal{G}}_{-}$, leading to, with respect to the convolution~\eqref{eq1.1}, the direct sum splitting $\tilde{\mathcal{G}}^{\ast}=\tilde{\mathcal{G}}_{+}^{\ast}\oplus\tilde{\mathcal{G}}_{-}^{\ast}$. Defining now the set $\mathrm{I}\big(\tilde{\mathcal{G}}^{\ast}\big)$ of Casimir invariant smooth functionals $h\colon \tilde{\mathcal{G}}^{\ast}\rightarrow \mathbb{R}$ on the adjoint space $\tilde{\mathcal{G}}^{\ast}$ via the coadjoint Lie algebra~$\tilde{\mathcal{G}}$ action
\begin{gather*}
\operatorname{ad}_{\nabla h\big(\tilde{l}\big)}^{\ast}\tilde{l}=0
\end{gather*}
at a seed element $\tilde{l}\in \tilde{\mathcal{G}}^{\ast}$, by means of the classical Adler--Kostant--Symes scheme~\cite{BlPrSa,Blas,FaTa,ReSe} one can generate \cite{BlHenPr,HePrBlPr,Ovsi-1,Ovsi-2,SeSz} a wide class of multi-dimensional completely integrable dispersionless (heavenly type) mutually commuting Hamiltonian systems
\begin{gather}
{\rm d}\tilde{l}/{\rm d}t:=-\operatorname{ad}_{\nabla h_{+} (\tilde{l} )}^{\ast}\tilde{l}, \label{eq1.3}
\end{gather}
for all $h\in \mathrm{I}\big(\tilde{\mathcal{G}}^{\ast}\big)$, $\nabla h\big(\tilde{l}\big):=\nabla h_{+}\big(\tilde{l}\big)\oplus\nabla h_{-}\big(\tilde{l}\big)\in\tilde{\mathcal{G}}_{+}\oplus\tilde{\mathcal{G}}_{-}$, on suitable functional manifolds. It is worth mentioning that the AKS scheme is a partial
case of the general $R$-matrix approach~\mbox{\cite{FaTa, ReSe}}, and which will be specified in Section~\ref{Sec_2}. Moreover, these mutually commuting flows~\eqref{eq1.3} can be equivalently represented as a commuting system of Lax--Sato type~\cite{HePrBlPr} vector field equations on the functional space $C^{2}(\mathbb{T}^{n}\times\mathbb{C};\mathbb{C})$ and generating a complete set of first integrals for them. As has been shown, amongst them there are
currently important equations for modern studies in physics, hydrodynamics and, in particular, in Riemannian geometry. Geometrically, they are related to
such interesting conformal structures on Riemannian metric spaces as Einstein and Einstein--Weyl metric equations, the first and second Pleba\'{n}ski
conformal metric equations, Dunajski metric equations, etc. It was observed that some of them are ge\-ne\-ra\-ted by seed elements $\tilde{l}\in \tilde{\mathcal{G}}^{\ast}$, meromorphic at some points of the complex plane~$\mathbb{C}$, requiring analysis that called for some modification of the theoretical foundations. Moreover, the general differential-geometric structure of seed elements, related to some conformal metric equations, proved to be invariant subject to the spatial dimension of the Riemannian
spaces under consideration and made it possible to describe them analytically. We analyze the Einstein--Weyl metric equation, the modified Einstein--Weyl
metric equation, the Dunajski heavenly equation system, the first and second conformal structure generating equations, the inverse first Shabat reduction
heavenly equation and the first and modified Pleba\'{n}ski heavenly equations, the Husain heavenly equation, the general Monge equation and their
multi-dimensional generalizations. We also construct superconformal analogs of the Whitham heavenly equation. These and related aspects of the integrable
multi-dimensional conformal metric equations, mentioned above, are studied and presented in the sequel.

\section[The Lie-algebraic structures and integrable Hamiltonian systems]{The Lie-algebraic structures\\ and integrable Hamiltonian systems}\label{Sec_2}

Consider the loop Lie algebra $\tilde{\mathcal{G}}$, described above. This Lie
algebra has elements representable as $a(x;\lambda):=\big\langle a(x;\lambda
),\frac{\partial}{\partial\mathrm{x}}\big\rangle =\sum\limits_{j=1}^{n}a_{j}(x;\lambda)\frac{\partial}{\partial x_{j}}+a_{0}(x;\lambda)\frac{\partial
}{\partial\lambda}\in\tilde{\mathcal{G}}$ for some holomorphic in
$\lambda\in\mathbb{D}_{\pm}^{1}$ vectors $a(x;\lambda)\in \mathbb{E}\times\mathbb{E}^{n}$ for all $x\in\mathbb{T}^{n}$, where
$\frac{\partial}{\partial\mathrm{x}}:= \big(\frac{\partial}{\partial\lambda
},\frac{\partial}{\partial x_{1}},\frac{\partial}{\partial x_{2}},\dots,\frac{\partial}{\partial x_{n}}\big)^{\top}$ is the generalized Euclidean
vector gradient with respect to the vector variable $\mathrm{x}:=(\lambda,x)\in \mathbb{C}\times\mathbb{T}^{n}$. As mentioned above, the
Lie algebra $\tilde{\mathcal{G}}$ naturally splits into the direct sum of two subalgebras
\begin{gather*}
\tilde{\mathcal{G}}=\tilde{\mathcal{G}}_{+}\oplus\tilde{\mathcal{G}}_{-},
\end{gather*}
allowing to introduce on it the classical $\mathcal{R}$-structure
\begin{gather*}
\big[\tilde{a},\tilde{b}\big]_{\mathcal{R}}:=\big[\mathcal{R}\tilde{a},\tilde {b}\big]+\big[\tilde{a},\mathcal{R}\tilde{b}\big]
\end{gather*}
for any $\tilde{a},\tilde{b}\in\tilde{\mathcal{G}}$, where
\begin{gather*}
\mathcal{R}:=(P_{+}-P_{-})/2,
\end{gather*}
and
\begin{gather*}
P_{\pm}\tilde{\mathcal{G}}:=\tilde{\mathcal{G}}_{\pm}\subset\mathcal{\tilde {G}}.
\end{gather*}
The space $\tilde{\mathcal{G}}^{\ast}\simeq\tilde{\Lambda}^{1}\big(\mathbb{T}^{n}\times\mathbb{C}\big)$, adjoint to the Lie algebra $\tilde{\mathcal{G}}$ of vector fields on $\mathbb{T}^{n}\times\mathbb{C}$, is functionally identified with $\tilde{\mathcal{G}}$ subject to the metric~\eqref{eq1.1}. Now for arbitrary $f,g\in\mathrm{D} \big(\tilde{\mathcal{G}}^{\ast}\big)$, one can determine two Lie--Poisson type brackets
\begin{gather*}
\{f,g\}:=\big(\tilde{l},\big[\nabla f\big(\tilde{l}\big),\nabla g\big(\tilde{l}\big)\big]\big)
\end{gather*}
and
\begin{gather}
\{f,g\}_{\mathcal{R}}:=\big(\tilde{l},\big[\nabla f\big(\tilde{l}\big),\nabla g\big(\tilde {l}\big)\big]_{\mathcal{R}}\big), \label{eq2.14}
\end{gather}
where at any seed element $\bar{l}\in\tilde{\mathcal{G}}^{\ast}$ the gradient element $\nabla f\big(\tilde{l}\big)$ and $\nabla g\big(\tilde{l}\big)\in\tilde{\mathcal{G}}$ are calculated with respect to the metric~\eqref{eq1.1}.

Now we assume that a smooth function $\gamma\in \mathrm{I}\big(\tilde{\mathcal{G}}^{\ast}\big)$ is a Casimir invariant, that is
\begin{gather}
\operatorname{ad}_{\nabla\gamma (\tilde{l})}^{\ast}\tilde{l}=0 \label{eq2.15}
\end{gather}
for a chosen seed element $\tilde{l}\in\tilde{\mathcal{G}}^{\ast}$. As the
coadjoint mapping $\operatorname{ad}_{\nabla f(\tilde{l})}^{\ast}\colon \tilde{\mathcal{G}}^{\ast
}\rightarrow\tilde{\mathcal{G}}^{\ast}$ for any $f\in\mathrm{D}\big(\tilde{\mathcal{G}}^{\ast}\big)$ can be rewritten in the reduced form
\begin{gather*}
\operatorname{ad}_{\nabla f(\tilde{l})}^{\ast}\big(\tilde{l}\big)=\left\langle \frac{\partial
}{\partial\mathrm{x}},\circ\nabla f(l)\right\rangle \bar{l}+\sum_{j=1}^{n}\left\langle \left\langle l,\frac{\partial}{\partial\mathrm{x}}\nabla
f(l)\right\rangle ,{\rm d}\mathrm{x}\right\rangle ,
\end{gather*}
where $\nabla f\big(\tilde{l}\big):=\big\langle \nabla f(l),\frac{\partial}{\partial\mathrm{x}}\big\rangle$. For the Casimir function $\gamma\in\mathrm{D}\big(\tilde{\mathcal{G}}^{\ast}\big)$ the condition~\eqref{eq2.15} is then equivalent to the equation
\begin{gather}
l\left\langle \frac{\partial}{\partial\mathrm{x}},\nabla\gamma(l)\right\rangle
+\left\langle \nabla\gamma(l),\frac{\partial}{\partial\mathrm{x}}\right\rangle
l+\left\langle l,\left(\frac{\partial}{\partial\mathrm{x}}\nabla\gamma (l)\right)\right\rangle =0, \label{eq2.18}
\end{gather}
which can be solved analytically. In the case when an element $\tilde{l}\in\tilde{\mathcal{G}}^{\ast}$ is singular as $|\lambda|\rightarrow\infty$, one can consider the general asymptotic expansion
\begin{gather}
\nabla\gamma:=\nabla\gamma^{(p)}\sim\lambda^{p}\sum\limits_{j\in \mathbb{Z}_{+}}\nabla\gamma_{j}^{(p)}\lambda^{-j} \label{eq2.19}
\end{gather}
for $p\in\mathbb{Z}_{+}$, and upon substituting \eqref{eq2.19} into the equation~\eqref{eq2.18}, one can solve it recurrently.

Now, let $h^{(y)},h^{(t)}\in \mathrm{I}\big(\tilde{\mathcal{G}}^{\ast}\big)$ be Casimir functions for which the Hamiltonian vector field ge\-nerators
\begin{gather}
\nabla h_{+}^{(y)}(l):=\big(\nabla\gamma^{(p_{y})}(l)\big)\big|_{+},\qquad \nabla h_{+}^{(t)}(l):=\big(\nabla h^{(p_{t})}(l)\big)\big|_{+} \label{eq2.20}
\end{gather}
are defined for specially chosen integers $p_{y},p_{t}\in\mathbb{Z}_{+}$. These invariants generate, owing to the Lie--Poisson bracket~\eqref{eq2.14}, the following commuting flows
\begin{gather}
\partial l/\partial t =-\left\langle \frac{\partial}{\partial
\mathrm{x}},\circ\nabla h_{+}^{(t)}(l)\right\rangle l-\left\langle
l,\left(\frac{\partial}{\partial\mathrm{x}}\nabla h_{+}^{(t)}(l)\right)\right\rangle \label{eq2.21a}
\end{gather}
and
\begin{gather}
\partial l/\partial y\text{ }=-\left\langle \frac{\partial}{\partial
\mathrm{x}},\circ\nabla h_{+}^{(y)}(l)\right\rangle l-\left\langle
l,\left(\frac{\partial}{\partial\mathrm{x}}\nabla h_{+}^{(y)}(l)\right)\right\rangle ,\label{eq2.21b}
\end{gather}
where $y,t\in\mathbb{R}$ are the corresponding evolution parameters. Since the invariants $h^{(y)},h^{(t)}\in \mathrm{I}\big(\tilde{\mathcal{G}}^{\ast}\big)$
commute with respect to the Lie--Poisson bracket~\eqref{eq2.14}, the flows~\eqref{eq2.21a} and~\eqref{eq2.21b} also commute, implying that the
corresponding Hamiltonian vector field generators
\begin{gather}
A_{\nabla h_{+}^{(t)}}:=\left\langle \nabla h_{+}^{(t)}(l),\frac{\partial
}{\partial\mathrm{x}}\right\rangle ,\qquad A_{\nabla h_{+}^{(y)}
}:=\left\langle \nabla h_{+}^{(y)}(l),\frac{\partial}{\partial\mathrm{x}}\right\rangle \label{eq2.22}
\end{gather}
satisfy the Lax compatibility condition
\begin{gather}
\frac{\partial}{\partial y}A_{\nabla h_{+}^{(t)}}-\frac{\partial}{\partial
t}A_{\nabla h_{+}^{(y)}}=\big[A_{\nabla h_{+}^{(t)}},A_{\nabla h_{+}^{(y)}}\big]\label{eq2.23}
\end{gather}
for all $y,t\in\mathbb{R}$. On the other hand, \eqref{eq2.23} is equivalent to the compatibility condition of two linear equations
\begin{gather}
\left( \frac{\partial}{\partial t}+A_{\nabla h_{+}^{(t)}}\right)
\psi=0,\qquad \left( \frac{\partial}{\partial y}+A_{\nabla h_{+}^{(y)}}\right) \psi=0 \label{eq2.24}
\end{gather}
for a function $\psi\in C^{2}\big(\mathbb{R}^{2}\times\mathbb{T}^{n} \times\mathbb{C};\mathbb{C}\big)$ for all $y,t\in\mathbb{R}$ and any $\lambda \in\mathbb{C}$. The above can be formulated as the following key result:

\begin{Proposition}\label{Prop_2.1} Let a seed vector field be $\tilde{l}\in\tilde{\mathcal{G}}^{\ast}$ and $h^{(y)},h^{(t)}\in \mathrm{I}\big(\tilde{\mathcal{G}}^{\ast}\big)$
be Casimir functions subject to the metric $(\cdot|\cdot)$ on the loop Lie algebra $\tilde{\mathcal{G}}$ and the natural coadjoint action on the loop co-algebra $\tilde{\mathcal{G}}^{\ast}$. Then the following dynamical systems
\begin{gather*}
\partial\tilde{l}/\partial y=-\operatorname{ad}_{\nabla h_{+}^{(y)}(\tilde{l})}^{\ast}
\tilde{l},\qquad \partial\tilde{l}/\partial t=-\operatorname{ad}_{\nabla h_{+}^{(t)}(\tilde{l})}^{\ast}\tilde{l}
\end{gather*}
are commuting Hamiltonian flows for all $y,t\in\mathbb{R}$. Moreover, the compatibility condition of these flows is equivalent to the vector field representations \eqref{eq2.24}, where $\psi\in C^{2}\big(\mathbb{R}^{2}\times\mathbb{T}^{n}\times\mathbb{C};\mathbb{C}\big)$ and the vector fields $A_{\nabla h_{+}^{(y)}},A_{\nabla h_{+}^{(t)}}\in\tilde{\mathcal{G}}$ are given by the expressions \eqref{eq2.22} and~\eqref{eq2.20}.
\end{Proposition}

\begin{Remark}\label{Rem2.2} As mentioned above, the expansion \eqref{eq2.19} is effective if a chosen seed element $\tilde{l}\in\tilde{\mathcal{G}}^{\ast}$ is singular
as $|\lambda|\rightarrow\infty$. In the case when it is singular as $|\lambda|\rightarrow0$, the expression \eqref{eq2.19} should be replaced by the expansion
\begin{gather*}
\nabla\gamma^{(p)}(l)\sim\lambda^{-p}\sum\limits_{j\in\mathbb{Z}_{+}}\nabla\gamma_{j}^{(p)}(l)\lambda^{j}
\end{gather*}
for suitably chosen integers $p\in\mathbb{Z}_{+}$, and the reduced Casimir function gradients then are given by the Hamiltonian vector field generators
\begin{align*}
& \nabla h_{-}^{(y)}(l):=\nabla\gamma^{(p_{y})}(l)|_{-},\\
& \nabla h_{-}^{(t)}(l):=\nabla\gamma^{(p_{t})}(l)|_{-}
\end{align*}
for suitably chosen positive integers $p_{y},p_{t}\in\mathbb{Z}_{+}$ and the corresponding Hamiltonian flows are written as $\partial\tilde{l}/\partial t=\operatorname{ad}_{\nabla h_{-}^{(t)} (\tilde{l} )}^{\ast}\tilde{l}$ and $\partial
\tilde{l}/\partial y=\operatorname{ad}_{\nabla h_{-}^{(y)} (\tilde{l} )}^{\ast}\tilde{l}$.
\end{Remark}

We add here for completeness that a seed element $\tilde{l}=\langle l,{\rm d}\mathrm{x} \rangle \in\tilde{\mathcal{G}}$ solves the Casimir equation~(\ref{eq2.18}),
generated by the chosen vector fields (\ref{eq2.22}), belonging by definition to a basis loop Lie algebra $\tilde{\mathcal{G}}_{\mathcal{+}}$ or $\tilde{\mathcal{G}}_{\mathcal{-}}$, making it possible to reconstruct the desired seed element, which is new and a~major advance over our previous work. As
for possible reductions, they can be \textit{a priori} imposed on the seed element via the related differential 1-form $\tilde{l}:=\langle l,{\rm d}\mathrm{x}\rangle \in\tilde{\mathcal{G}}^{\ast}$ on the set $\mathbb{T}^{n}\times\mathbb{C}$ in most studied cases satisfying the following conditions: $\tilde{l}=\eta {\rm d}\rho$ for some mappings $\eta,\rho\colon \mathbb{T}^{n}\times\mathbb{C}\rightarrow \mathbb{C}$, or ${\rm d}\tilde{l}=0$; that is, $d\eta\wedge {\rm d}\rho=0$ on $\mathbb{T}^{n}\times\mathbb{C}$.

It is also worth of mentioning that, following Ovsienko's scheme \cite{Ovsi-1,Ovsi-2}, one can consider a~slightly wider class of integrable heavenly equations, realized as compatible Hamiltonian flows on the semidirect product of the holomorphic loop Lie algebra $\tilde{\mathcal{G}}$ of vector fields on $\mathbb{T}^{n}\times\mathbb{C}$ and its regular coadjoint space $\tilde{\mathcal{G}}^{\ast}$, supplemented with naturally related cocycles.

\section[Lax--Sato integrable multi-dimensional heavenly systems and related
conformal structure generating equations]{Lax--Sato integrable multi-dimensional heavenly systems\\ and related
conformal structure generating equations}\label{Sec 3}

In this section we apply our method to several examples.

\subsection{Einstein--Weyl metric equation}

Define $\tilde{\mathcal{G}}^{\ast}=\widetilde{{\rm diff}}\big(\mathbb{T}^{1} \times\mathbb{C}\big)^{\ast}$ and take the seed element
\begin{gather*}
\tilde{l}=\big( u_{x}\lambda-2u_{x}v_{x}-u_{y}\big) {\rm d}x+\big(\lambda^{2}-v_{x}\lambda+v_{y}+{v_{x}^{2}}\big) {\rm d}\lambda,
\end{gather*}
which generates with respect to the metric \eqref{eq1.1} the gradient of the Casimir invariants $h^{(p_{t})},h^{(p_{y})}\allowbreak \in \mathrm{I}\big(\tilde {\mathcal{G}}^{\ast}\big)$ in the form
\begin{gather*}
 \nabla h^{(p_{t})}(l)\sim\lambda^{2}(0,1)^{\top}+(-u_{x},v_{x})^{\top}\lambda+(u_{y},u-v_{y})^{\top}+O\big(\lambda^{-1}\big),\\
 \nabla h^{(p_{y})}(l)\sim\lambda(0,1)^{\top}+(-u_{x},v_{x})^{\top} +(u_{y},-v_{y})^{\top}\lambda^{-1}+O\big(\lambda^{-2}\big)
\end{gather*}
as $|\lambda|\rightarrow\infty$ at $p_{t}=2$, $p_{y}=1$. For the gradients of the Casimir functions $h^{(t)},h^{(y)}\in \mathrm{I}\big(\tilde{\mathcal{G}}^{\ast}\big)$ determined by \eqref{eq2.20}, one can easily obtain the corresponding Hamiltonian vector field generators
\begin{gather}
A_{\nabla h_{+}^{(t)}}=\left\langle \nabla h_{+}^{(t)}(l),\frac{\partial
}{\partial\mathrm{x}}\right\rangle =\big(\lambda^{2}+\lambda v_{x}+u-v_{y} \big)\frac{\partial}{\partial x}+(-\lambda u_{x}+u_{y})\frac{\partial}
{\partial\lambda},\nonumber\\
A_{\nabla h_{+}^{(y)}}=\left\langle \nabla h_{+}^{(y)}(l),\frac{\partial
}{\partial\mathrm{x}}\right\rangle =(\lambda+v_{x})\frac{\partial}{\partial
x}-u_{x}\frac{\partial}{\partial\lambda}, \label{eq2.27}
\end{gather}
satisfying the compatibility condition \eqref{eq2.23}, which is equivalent to the set of equations
\begin{gather}
u_{xt}+u_{yy}+(uu_{x})_{x}+v_{x}u_{xy}-v_{y}u_{xx}=0,\nonumber\\
v_{xt}+v_{yy}+uv_{xx}+v_{x}v_{xy}-v_{y}v_{xx}=0, \label{eq2.28}
\end{gather}
describing general integrable Einstein--Weyl metric equations~\cite{DuMaTo}.

As is well known \cite{MaSa}, the invariant reduction of \eqref{eq2.28} at $v=0$ gives rise to the famous dispersionless Kadomtsev--Petviashvili equation
\begin{gather}
(u_{t}+uu_{x})_{x}+u_{yy}=0, \label{eq2.29}
\end{gather}
for which the reduced vector field representation \eqref{eq2.24} follows from \eqref{eq2.27} and is given by the vector fields
\begin{gather}
 A_{\nabla h_{+}^{(t)}}=\big(\lambda^{2}+u\big)\frac{\partial}{\partial x}+(-\lambda
u_{x}+u_{y})\frac{\partial}{\partial\lambda},\nonumber\\
 A_{\nabla h_{+}^{(y)}}=\lambda\frac{\partial}{\partial x}-u_{x}\frac{\partial}{\partial\lambda}, \label{eq2.30}
\end{gather}
satisfying the compatibility condition \eqref{eq2.23}, which is equivalent to the equation \eqref{eq2.29}. In particular, one finds from~\eqref{eq2.24} and~\eqref{eq2.30} that the vector field compatibility relationships
\begin{gather*}
 \frac{\partial\psi}{\partial t}+\big(\lambda^{2}+u\big)\frac{\partial\psi}{\partial
x}+(-\lambda u_{x}+u_{y})\frac{\partial\psi}{\partial\lambda}=0,\\
 \frac{\partial\psi}{\partial y}+\lambda\frac{\partial\psi}{\partial
x}-u_{x}\frac{\partial\psi}{\partial\lambda}=0,
\end{gather*}
are satisfied for $\psi\in C^{2}\big(\mathbb{R}^{2}\times\mathbb{T}^{1}\times\mathbb{C};\mathbb{C}\big)$ and any $y,t\in\mathbb{R}$, $(x,\lambda)\in\mathbb{T}^{1}\times\mathbb{C}$.

\subsection{The modified Einstein--Weyl metric equation}

This equation system is
\begin{gather*}
 u_{xt}=u_{yy}+u_{x}u_{y}+u_{x}^{2}w_{x}+uu_{xy}+u_{xy}w_{x}+u_{xx}a,\\
 w_{xt}=uw_{xy}+u_{y}w_{x}+w_{x}w_{xy}+aw_{xx}-a_{y},
\end{gather*}
where $a_{x}:=u_{x}w_{x}-w_{xy}$, as was recently derived in \cite{Szab}. In this case we take also $\tilde{\mathcal{G}}^{\ast}=\widetilde{{\rm diff}}
\big(\mathbb{T}^{1}\times\mathbb{C}\big)$, yet for a seed element $\tilde{l}\in \tilde{\mathcal{G}}$ we choose the form
\begin{gather*}
 \tilde{l}=\big[{{\lambda}^{2}}u_{x}+ ( 2u_{x}w_{x}+u_{y}+3{u}u_{x} )
\lambda+2u_{x}\partial_{x}^{-1}{u_{x}w_{x} }+2u_{x}
\partial_{x}^{-1}{ u_{y} }\\
\hphantom{\tilde{l}=}{}+3u_{x}{{w_{x}}^{2}}+2u_{y}w_{x}+6{u}u_{x}w_{x}+2{u}u_{y}+3{{u}^{2}}u_{x}-2{a}u_{x}\big]{\rm d}x \\
\hphantom{\tilde{l}=}{} +\big[{{\lambda}^{2}}+ ( w_{x}+3{u} ) \lambda+2\partial
_{x}^{-1}{ u_{x}w_{x} }+2\partial_{x}^{-1}{u_{y}}+{{w_{x}}^{2}}+3{u}w_{x}+3{{u}^{2}}-{a}\big]{\rm d}\lambda,
\end{gather*}
which with respect to the metric \eqref{eq1.1} generates two Casimir invariants $\gamma^{(j)}\in \mathrm{I}\big(\tilde{\mathcal{G}}^{\ast}\big)$, $j= 1,2$, whose gradients are
\begin{gather*}
 \nabla\gamma^{(1)}(l)\sim\lambda\big[(u_{x},-1)^{\top}+(0,w_{x})^{\top }\lambda^{-1}\big]+O\big(\lambda^{-1}\big),\\
 \nabla\gamma^{(2)}(l)\sim\lambda^{2}\big[(u_{x},-1)^{\top}+(uu_{x}+u_{y},-u+w_{x})^{\top}\lambda^{-1}+ (0,uw_{x}-a)^{\top}\lambda^{-2}\big]+O\big(\lambda^{-1}\big),
\end{gather*}
as $|\lambda|\rightarrow\infty$ at $p_{y}=1$, $p_{t}=2$. The corresponding gradients of the Casimir functions $h^{(t)},h^{(y)}\in \mathrm{I} \big(\mathcal{G}^{\ast}\big)$, determined by~\eqref{eq2.20}, generate the Hamiltonian vector field expressions
\begin{gather}
 \nabla h_{+}^{(y)}:=\nabla\gamma^{(1)}(l)|_{+}=(u_{x}\lambda,-\lambda +w_{x})^{\top},\nonumber\\
\nabla h_{+}^{(t)}=\nabla\gamma^{(2)}(l)|_{+}=\big(u_{x}\lambda^{2}+(uu_{x}+u_{y})\lambda,-\lambda^{2}+(w_{x}-u)\lambda+uw_{x}-a\big)^{\top}.\label{eq2.31c}
\end{gather}
Now one easily obtains from \eqref{eq2.31c} the compatible Lax system of linear equations
\begin{gather*}
 \frac{\partial\psi}{\partial y}+(-\lambda+w_{x})\frac{\partial\psi }{\partial x}+u_{x}\lambda\frac{\partial\psi}{\partial\lambda}=0,\\
\frac{\partial\psi}{\partial t}+\big({-}\lambda^{2}+(w_{x}-u)\lambda +uw_{x}-a\big)\frac{\partial\psi}{\partial x}+\big(u_{x}\lambda^{2}+(uu_{x}+u_{y}\big)\lambda)\frac{\partial\psi}{\partial\lambda}=0,
\end{gather*}
satisfied for $\psi\in C^{2}\big(\mathbb{R}^{2}\times\mathbb{T}^{1}\times \mathbb{C};\mathbb{C}\big)$ and any $y,t\in\mathbb{R}$, $(x,\lambda)\in\mathbb{T}^{1}\times\mathbb{C}$.

\subsection{The Dunajski heavenly equation system}

This equation, suggested in \cite{Duna}, generalizes the corresponding anti-self-dual vacuum Einstein equation with zero cosmological constant, which is related to the Pleba\'{n}ski metric and the ce\-lebrated Pleba\'{n}ski~\cite{FeKr,Pleb} second heavenly equation. To study the integrability of the Dunajski equations
\begin{gather}
u_{x_{1}t}+u_{yx_{2}}+u_{x_{1}x_{1}}u_{x_{2}x_{2}}-u_{x_{1}x_{2}}^{2}-v=0,\nonumber\\
v_{x_{1}t}+v_{x_{2}y}+u_{x_{1}x_{1}}v_{x_{2}x_{2}}-2u_{x_{1}x_{2}}v_{x_{1}x_{2}}=0, \label{eq2.44}%
\end{gather}
where $(u,v)\in C^{\infty}\big(\mathbb{R}^{2}\times\mathbb{T}^{2}\times \mathbb{C};\mathbb{R}^{2}\big)$, $(y,t;x_{1},x_{2})\in\mathbb{R}^{2}\times\mathbb{T}_{\mathbb{C}}^{2}$, we define $\tilde{\mathcal{G}}^{\ast }=\widetilde{{\rm diff}}\big(\mathbb{T}^{2}\times\mathbb{C}\big)^{\ast}$ and choose the following seed element $\bar{l}\in\tilde{\mathcal{G}}^{\ast}$:
\begin{gather*}
\tilde{l}=(\lambda+v_{x_{1}}-u_{x_{1}x_{1}}+u_{x_{1}x_{2}}){\rm d}x_{1} +(\lambda+v_{x_{2}}+u_{x_{2}x_{2}}-u_{x_{1}x_{2}}){\rm d}x_{2}+(\lambda-x_{1}-x_{2}){\rm d}\lambda.
\end{gather*}
With respect to the metric~\eqref{eq1.1}, the gradients of two functionally independent Casimir invariants $h^{(p_{y})},h^{(p_{y})}\in \mathrm{I}\big(\tilde{\mathcal{G}}^{\ast}\big)$ can be obtained as $|\lambda|\rightarrow \infty$ in the asymptotic form as
\begin{gather}
 \nabla h^{(p_{y})}(l)\sim\lambda(1,0,0)^{\top}+(-u_{x_{1}x_{2}},u_{x_{1}x_{1}},-v_{x_{1}})^{\top}+O\big(\lambda^{-1}\big),\nonumber\\
\nabla h^{(p_{t})}(l)\sim\lambda(0,-1,0)^{\top}+(u_{x_{2}x_{2}},-u_{x_{1}x_{2}},v_{x_{2}})^{\top}+O\big(\lambda^{-1}\big), \label{eq2.46}%
\end{gather}
at $p_{t}=1=p_{y}$. Upon calculating the Hamiltonian vector field generators
\begin{gather*}
 \nabla h_{+}^{(y)}:=\nabla h^{(p_{y})}(l)|_{+}=(\lambda-u_{x_{1}x_{2}},u_{x_{1}x_{1}},-v_{x_{1}})^{\top},\\
\nabla h_{+}^{(t)}:=\nabla h^{(p_{t})}(l)|_{+}=(u_{x_{2}x_{2}},-\lambda-u_{x_{1}x_{2}},v_{x_{2}})^{\top},
\end{gather*}
following from the Casimir functions gradients \eqref{eq2.46}, one readily obtains the following vector fields
\begin{gather*}
A_{\nabla h_{+}^{(t)}}=\left\langle \nabla h_{+}^{(t)},\frac{\partial}{\partial \mathrm{x}}\right\rangle=u_{x_{2}x_{2}}\frac{\partial}{\partial x_{1}}-(\lambda+u_{x_{1}x_{2}})\frac{\partial}{\partial x_{2}}+v_{x_{2}}\frac{\partial}{\partial\lambda},\\
A_{\nabla h_{+}^{(y)}}=\left\langle \nabla h_{+}^{(y)},\frac{\partial}{\partial\mathrm{x}}\right\rangle =(\lambda-u_{x_{1}x_{2}})\frac{\partial}{\partial x_{1}}+u_{x_{1}x_{1}}\frac{\partial}{\partial x_{2}}-v_{x_{1}}\frac{\partial}{\partial\lambda},
\end{gather*}
satisfying the Lax compatibility condition \eqref{eq2.23}, which is equivalent to the vector field compatibility relationships
\begin{gather*}
\frac{\partial\psi}{\partial t}+u_{x_{2}x_{2}}\frac{\partial\psi}{\partial x_{1}}-(\lambda+u_{x_{1}x_{2})}\frac{\partial\psi}{\partial x_{2}}+v_{x_{2}}\frac{\partial\psi}{\partial\lambda}=0,\\
\frac{\partial\psi}{\partial y}+(\lambda-u_{x_{1}x_{2}})\frac{\partial\psi}{\partial x_{1}}+u_{x_{1}x_{1}}\frac{\partial\psi}{\partial x_{2}}-v_{x_{1}}\frac{\partial\psi}{\partial\lambda}=0,
\end{gather*}
satisfied for $\psi\in C^{2}\big(\mathbb{R}^{2}\times\mathbb{T}^{2}\times \mathbb{C};\mathbb{C}\big)$, any $(y,t)\in\mathbb{R}^{2}$ and all $(x_{1},x_{2};\lambda)\in\mathbb{T}^{2}\times\mathbb{C}$. As mentioned in~\cite{BoDrMa}, the Dunajski equations~\eqref{eq2.44} generalize both the dispersionless Kadomtsev--Petviashvili and Pleba\'{n}ski second heavenly equations, and is also a Lax integrable Hamiltonian system.

\subsection[First conformal structure generating equation: $u_{yt}+u_{xt}u_{y}-u_{t}u_{xy}=0$]{First conformal structure generating equation: $\boldsymbol{u_{yt}+u_{xt}u_{y}-u_{t}u_{xy}=0}$}

The seed element $\tilde{l}\in\tilde{\mathcal{G}}^{\ast}\simeq\widetilde{{\rm diff}}\big(\mathbb{T}^{1}\times\mathbb{C}\big)^{\ast}$ in the form
\begin{gather*}
\tilde{l}=\big[u_{t}^{-2}(1-\lambda)\lambda^{-1}+u_{y}^{-2}\lambda(\lambda -1)^{-1}\big]{\rm d}x,
\end{gather*}
where $u\in C^{2}\big(\mathbb{R}^{2}\times\mathbb{T}^{1}\times\mathbb{C};\mathbb{R}\big)$, $x\in\mathbb{T}^{n}$, $\lambda\in\mathbb{C}\backslash\{0,1\}$, generates two independent Casimir functionals $\gamma^{(1)}$ and $\gamma^{(2)}\in \mathrm{I}\big(\tilde{\mathcal{G}}^{\ast}\big)$, having gradients with the asymptotic expansions
\begin{gather*}
\nabla\gamma^{(1)}(l)\sim u_{y}+O\big(\mu^{2}\big),
\end{gather*}
as $|\mu|\rightarrow0$, $\mu:=\lambda-1$, and
\begin{gather*}
\nabla\gamma^{(2)}(l)\sim u_{t}+O\big(\lambda^{2}\big),
\end{gather*}
as $|\lambda|\rightarrow0$. The commutativity condition
\begin{gather}
\big[ X^{(y)},X^{(t)}\big]=0 \label{C2}
\end{gather}
of the vector fields
\begin{gather}
X^{(y)}:=\partial/\partial y+\nabla h^{(y)}\big(\tilde{l}\big),\qquad X^{(t)}:=\partial/\partial t+\nabla h^{(t)}\big(\tilde{l}\big), \label{C2a}%
\end{gather}
where
\begin{gather*}
\nabla h^{(y)}\big(\tilde{l}\big):=-\big(\mu^{-1}\nabla\gamma^{(1)}\big(\tilde{l}\big)\big)\big|_{-}=-\frac{u_{y}}{\lambda-1}\frac{\partial}{\partial x},\\
\nabla h^{(t)}\big(\tilde{l}\big):=-\big(\lambda^{-1}\nabla\gamma^{(2)}\big(\tilde{l}\big)\big)\big|_{-}=-\frac{u_{t}}{\lambda}\frac{\partial}{\partial x},
\end{gather*}
leads to the heavenly type equation
\begin{gather*}
u_{yt}+u_{xt}u_{y}-u_{xy}u_{t}=0,
\end{gather*}
considered in~\cite{AdSh}. Its Lax--Sato representation is the compatibility condition for the first-order partial differential equations
\begin{gather*}
\frac{\partial\psi}{\partial y}-\frac{u_{y}}{\lambda-1}\frac{\partial\psi}{\partial x}=0,\\
\frac{\partial\psi}{\partial t}-\frac{u_{t}}{\lambda}\frac{\partial\psi}{\partial x}=0,
\end{gather*}
where $\psi\in C^{2}\big(\mathbb{R}^{2}\times\mathbb{T}^{1}\times\mathbb{C};\mathbb{C}\big)$.

\subsection[Second conformal structure generating equation: $u_{xt} + u_{x}u_{yy}-u_{y}u_{xy}=0$]{Second conformal structure generating equation:\\ $\boldsymbol{u_{xt} + u_{x}u_{yy}-u_{y}u_{xy}=0}$}

For a seed element $\tilde{l}\in\tilde{\mathcal{G}}^{\ast}=\widetilde{{\rm diff}}\big(\mathbb{T}^{1}\times\mathbb{C}\big)^{\ast}$ in the form
\begin{gather*}
\tilde{l}=\big[u_{x}^{2}+2u_{x}^{2}(u_{y}+\alpha)\lambda^{-1}+u_{x}^{2}\big(3u_{y}^{2}+4\alpha u_{y}+\beta\big)\lambda^{-2}\big]{\rm d}x,
\end{gather*}
where $u\in C^{2}\big(\mathbb{R}^{2}\times\mathbb{T}^{1}\times\mathbb{C};\mathbb{R}\big)$, $x\in\mathbb{T}^{1}$, $\lambda\in\mathbb{C}\setminus\{0\}$, and $\alpha,\beta\in\mathbb{R}$, there is one independent Casimir functional $\gamma\in \mathrm{I}\big(\tilde{\mathcal{G}}^{\ast}\big)$ with the following asymptotic as $|\lambda|\rightarrow0$ expansion of its functional gradient
\begin{gather*}
\nabla\gamma^{(1)}(l)\sim c_{0}u_{x}^{-1}+(-c_{0}u_{y}+c_{1})u_{x}^{-1}\lambda+(-c_{1}u_{y}+c_{2})u_{x}^{-1}\lambda^{2}+O\big(\lambda^{3}\big),
\end{gather*}
where $c_{r}\in\mathbb{R}$, $r=1,2$. Assuming that $c_{0}=1$, $c_{1}=0$ and $c_{2}=0$, we obtain two functionally independent gradient elements
\begin{gather*}
\nabla h^{(y)}\big(\tilde{l}\big):=-\big(\lambda^{-1}\nabla\gamma^{(1)}\big(\tilde{l}\big)\big)\big|_{-}=-\frac{1}{\lambda u_{x}}\frac{\partial}{\partial x},\\
\nabla h^{(t)}\big(\tilde{l}\big):=\big(\lambda^{-2}\nabla\gamma^{(1)}\big(\tilde{l}\big)\big)\big|_{-}=\left( \frac{1}{\lambda^{2}u_{x}}-\frac{u_{y}}{\lambda u_{x}}\right)
\frac{\partial}{\partial x}.
\end{gather*}
The corresponding commutativity condition \eqref{C2} of the vector fields~\eqref{C2a} give rise to the heavenly equation
\begin{gather*}
u_{xt}+u_{x}u_{yy}-u_{y}u_{xy}=0,
\end{gather*}
whose linearized Lax--Sato representation is given by the first-order system
\begin{gather*}
\frac{\partial\psi}{\partial y}-\frac{1}{\lambda u_{x}}\frac{\partial\psi}{\partial x}=0,\\
\frac{\partial\psi}{\partial t}+\left( \frac{1}{\lambda^{2}u_{x}}-\frac{u_{y}}{\lambda u_{x}}\right) \frac{\partial\psi}{\partial x}=0,
\end{gather*}
of linear vector field equations for a function $\psi\in C^{2}\big(\mathbb{R}^{2}\times\mathbb{T}^{1}\times\mathbb{C};\mathbb{C}\big)$.

\subsection{Inverse first Shabat reduction heavenly equation}

The seed element $\tilde{l}\in\tilde{\mathcal{G}}^{\ast}=\widetilde{{\rm diff}}\big(\mathbb{T}^{1}\times\mathbb{C}\big)^{\ast}$ in the form
\begin{gather*}
\tilde{l}=\big(a_{0}u_{y}^{-2}u_{x}^{2}(\lambda+1)^{-1}+a_{1}u_{x}^{2}+a_{1}u_{x}^{2}\lambda\big){\rm d}x,
\end{gather*}
where $u\in C^{2}\big(\mathbb{R}^{2}\times\mathbb{T}^{1}\times\mathbb{C};\mathbb{R}\big)$, $x\in\mathbb{T}^{1}$, $\lambda\in\mathbb{C}\setminus\{-1\}$, and $a_{0},a_{1}\in\mathbb{R}\backslash\{0\}$, generates two independent Casimir functionals $\gamma^{(1)}$ and $\gamma^{(2)}\in \mathrm{I}\big(\tilde{\mathcal{G}}^{\ast}\big)$, whose gradients have the following asymp\-to\-tic expansions
\begin{gather*}
\nabla\gamma^{(1)}(l)\sim u_{y}u_{x}^{-1}-u_{y}u_{x}^{-1}\mu+O\big(\mu^{2}\big),
\end{gather*}
as $|\mu|\rightarrow0$, $\mu:=\lambda+1$, and
\begin{gather*}
\nabla\gamma^{(2)}(l)\sim u_{x}^{-1}+O\big(\lambda^{-2}\big),
\end{gather*}
as $|\lambda|\rightarrow\infty$. If we define
\begin{gather*}
\nabla h^{(y)}\big(\tilde{l}\big):=\big(\mu^{-1}\nabla\gamma^{(1)}\big(\tilde{l}\big)\big)\big|_{-}=-\frac{\lambda}{\lambda+1}\frac{u_{y}}{u_{x}}\frac{\partial}{\partial x},\\
\nabla h^{(t)}\big(\tilde{l}\big):=\big(\lambda\nabla\gamma^{(2)}\big(\tilde{l}\big)\big)\big|_{+}=\frac{\lambda}{u_{x}}\frac{\partial}{\partial x},
\end{gather*}
the commutativity condition \eqref{C2} of the vector fields \eqref{C2a} leads to the heavenly equation
\begin{gather*}
u_{xy}+u_{y}u_{tx}-u_{ty}u_{x}=0.
\end{gather*}
We note that this can be obtained as a result of the simultaneously changing the independent variables $\mathbb{R}\ni x \rightarrow t\in\mathbb{R}$, $\mathbb{R} \ni y \rightarrow x\in\mathbb{R}$ and $\mathbb{R}\ni t \rightarrow y\in\mathbb{R}$ in the first Shabat reduction heavenly equation~\cite{AlSha}. The corresponding Lax--Sato representation is given by the compatibility condition for the first-order vector field equations
\begin{gather*}
\frac{\partial\psi}{\partial y}-\frac{\lambda}{\lambda+1}\frac{u_{y}}{u_{x}}\frac{\partial\psi}{\partial x}=0,\\
\frac{\partial\psi}{\partial t}+\frac{\lambda}{u_{x}}\frac{\partial\psi }{\partial x}=0,
\end{gather*}
where $\psi\in C^{2}\big(\mathbb{R}^{2}\times\mathbb{T}^{1}\times\mathbb{C};\mathbb{C}\big)$.

\subsection{First Pleba\'{n}ski heavenly equation and its generalizations}

We choose the seed element $\tilde{l}\in\tilde{\mathcal{G}}^{\ast}=\widetilde{{\rm diff}}\big(\mathbb{T}^{2}\times\mathbb{C}\big)^{\ast}$ in the form
\begin{gather*}
\tilde{l}=\big[u_{x_{1}x_{1}}-u_{x_{1}x_{2}}+\lambda(x_{1}+x_{2})+\lambda ^{-1}(u_{yx_{1}}+u_{tx_{1}})\big]{\rm d}x_{1}\\
\hphantom{\tilde{l}=}{} +\big[u_{x_{1}x_{2}}-u_{x_{2}x_{2}}+\lambda(x_{1}+x_{2})+\lambda^{-1}(u_{yx_{2}}+u_{tx_{2}})\big]{\rm d}x_{2}, %\label{H1}%
\end{gather*}
or
\begin{gather*}
\tilde{l}={\rm d}\big[u_{x_{1}}-u_{x_{2}}+\lambda(x_{1}+x_{2})+\lambda^{-1}(u_{y}+u_{t})\big], %\label{H1a}%
\end{gather*}
where ${\rm d}\lambda=0$, $u\in C^{2}\big(\mathbb{R}^{2}\times\mathbb{T}^{2};\mathbb{R}\big)$, $(x_{1},x_{2})\in\mathbb{T}^{2}$, $\lambda\in\mathbb{C}\backslash\{0\}$, generates two independent Casimir functionals $\gamma^{(1)}$ and $\gamma^{(2)}\in \mathrm{I}\big(\tilde{\mathcal{G}}^{\ast}\big)$, whose gradients have the following asymptotic expansions
\begin{gather*}
 \nabla\gamma^{(1)}(l)\sim(-u_{yx_{2}},u_{yx_{1}})^{\top}+O\big(\lambda^{2}\big),\nonumber\\
 \nabla\gamma^{(2)}(l)\sim(-u_{tx_{2}},u_{tx_{1}})^{\top}+O\big(\lambda^{2}\big),%\label{Hen*}%
\end{gather*}
as $|\lambda|\rightarrow0$. The commutativity condition \eqref{C2} of the vector fields~\eqref{C2a}, where
\begin{gather*}
\nabla h^{(y)}\big(\tilde{l}\big):=\big(\lambda^{-1}\nabla\gamma^{(1)}\big(\tilde{l}\big)\big)\big|_{-}=-\frac{u_{yx_{2}}}{\lambda}\frac{\partial}{\partial x_{1}}+\frac{u_{yx_{1}}}{\lambda}\frac{\partial}{\partial x_{2}},\\
\nabla h^{(t)}\big(\tilde{l}\big):=\big(\lambda^{-1}\nabla\gamma^{(2)}\big(\tilde{l}\big)\big)\big|_{-}=-\frac{u_{tx_{2}}}{\lambda}\frac{\partial}{\partial x_{1}} +\frac{u_{tx_{1}}}{\lambda}\frac{\partial}{\partial x_{2}},
\end{gather*}
leads to the first Pleba\'{n}ski heavenly equation~\cite{DoFe,ShYa3}
\begin{gather*}
u_{yx_{1}}u_{tx_{2}}-u_{yx_{2}}u_{tx_{1}}=1.
\end{gather*}
Its Lax--Sato representation entails the compatibility condition for the first-order partial dif\-fe\-rential equations{\samepage
\begin{gather*}
\frac{\partial\psi }{\partial y}-\frac{u_{yx_{2}}}{\lambda}\frac{\partial\psi }{\partial x_{1}}+\frac{u_{yx_{1}}}{\lambda}\frac{\partial\psi }{\partial x_{2}}=0,\\
\frac{\partial\psi }{\partial t}-\frac{u_{tx_{2}}}{\lambda}\frac{\partial\psi }{\partial x_{1}}+\frac{u_{tx_{1}}}{\lambda}\frac{\partial\psi }{\partial x_{2}}=0,
\end{gather*}
where $\psi \in C^{\infty}\big(\mathbb{R}^{2}\times\mathbb{T}^{2}\times\mathbb{C};\mathbb{C}\big)$.}

The proposed Lie-algebraic scheme can be easily generalized for any dimension $n=2k$, where $k\in\mathbb{N}$, and $n>2$. In this case one has $2k$ independent Casimir functionals $\gamma^{(j)}\in \mathrm{I}\big(\tilde {\mathcal{G}}^{\ast}\big)$, $j=1,\dots,2k$, with the following asymptotic expansions for their gradients:
\begin{gather*}
\nabla\gamma^{(1)}(l)\sim(-u_{yx_{2}},u_{yx_{1}},\underbrace{0,\ldots,0}_{2k-2})^{\top}+O\big(\lambda^{2}\big),\\
\nabla\gamma^{(2)}(l)\sim(-u_{tx_{2}},u_{tx_{1}},\underbrace{0,\ldots,0}_{2k-2})^{\top}+O\big(\lambda^{2}\big),\\
\cdots \cdots\cdots\cdots\cdots\cdots\cdots\cdots\cdots\cdots\cdots\cdots\cdots\cdots\\
\nabla\gamma^{(2k-1)}(l)\sim(\underbrace{0,\ldots,0}_{2k-2},-u_{yx_{2k}},u_{yx_{2k-1}})^{\top}+O\big(\lambda^{2}\big),\\
\nabla\gamma^{(2k)}(l)\sim(\underbrace{0,\ldots,0}_{2k-2},-u_{tx_{2k}},u_{tx_{2k-1}})^{\top}+O\big(\lambda^{2}\big),
\end{gather*}
as $|\lambda|\rightarrow0$, where $u\in C^{2}\big(\mathbb{R}^{2}\times \mathbb{T}^{2k};\mathbb{R}\big)$. If we set
\begin{gather*}
 \nabla h^{(y)}\big(\tilde{l}\big):=\big(\lambda^{-1}\big(\nabla\gamma^{(1)}\big(\tilde{l}\big)+\dots+\nabla\gamma^{(2k-1)}\big(\tilde{l}\big)\big)\big)\big|_{-} \\
\hphantom{\nabla h^{(y)}\big(\tilde{l}\big)}{}=-\sum_{m=1}^{k}\left( \frac{u_{yx_{2m}}}{\lambda}\frac{\partial}{\partial x_{2m-1}}-\frac{u_{yx_{2m-1}}}{\lambda}\frac{\partial}{\partial x_{2m}}\right) ,\\
 \nabla h^{(t)}\big(\tilde{l}\big):=\big(\lambda^{-1}\big(\nabla\gamma^{(2)}\big(\tilde{l}\big)+\dots+\nabla\gamma^{(2k)}\big(\tilde{l}\big)\big)\big)|_{-} \\
\hphantom{\nabla h^{(y)}\big(\tilde{l}\big)}{}=-\sum_{m=1}^{k}\left( \frac{u_{tx_{2m}}}{\lambda}\frac{\partial }{\partial x_{2m-1}}-\frac{u_{tx_{2m-1}}}{\lambda}\frac{\partial}{\partial x_{2m}}\right) ,
\end{gather*}
the commutativity condition \eqref{C2} of the vector fields \eqref{C2a} leads to the following multi-di\-men\-sio\-nal analogs of the first Pleba\'{n}ski heavenly equation
\begin{gather*}
\sum_{m=1}^{k}(u_{yx_{2m-1}}u_{tx_{2m}}-u_{yx_{2m}}u_{tx_{2m-1}})=1,
\end{gather*}
having the generating seed element
\begin{gather*}
\tilde{l}={\rm d}\big[u_{x_{1}}-u_{x_{2}}+\lambda(x_{1}+x_{2})+\lambda^{-1}(u_{y}+u_{t})\big], %\label{Hen1}
\end{gather*}
where $d\lambda=0$ and $u\in C^{2}\big(\mathbb{R}^{2}\times\mathbb{T}^{2k};\mathbb{R}\big)$.

\subsection{Modified Pleba\'{n}ski heavenly equation and its generalizations}

For the seed element $\tilde{l}\in\tilde{\mathcal{G}}^{\ast}=\widetilde{{\rm diff}}\big(\mathbb{T}^{2}\times\mathbb{C}\big)^{\ast}$ in the form
\begin{gather}
 \tilde{l}=\big(\lambda^{-1}u_{x_{1}y}+u_{x_{1}x_{1}}-u_{x_{1}x_{2}}+\lambda\big){\rm d}x_{1}+\big(\lambda^{-1}u_{x_{2}y}+u_{x_{1}x_{2}}-u_{x_{2}x_{2}}+\lambda\big){\rm d}x_{2} \nonumber\\
\hphantom{\tilde{l}}{} ={\rm d}(\lambda^{-1}u_{y}+u_{x_{1}}-u_{x_{2}}+\lambda x_{1}+\lambda x_{2}). \label{H3}%
\end{gather}
where $d\lambda=0$, $u\in C^{2}\big(\mathbb{R}^{2}\times\mathbb{T}^{2};\mathbb{R}\big)$, $(x_{1},x_{2})\in\mathbb{T}^{2}$, $\lambda\in\mathbb{C}\backslash\{0\}$, there exist two independent Casimir functionals $\gamma^{(1)}$ and $\gamma^{(2)}\in \mathrm{I}\big(\tilde{\mathcal{G}}^{\ast}\big)$ with the gradient asymptotic expansions
\begin{gather*}
\nabla\gamma^{(1)}(l)\sim(u_{yx_{2}},-u_{yx_{1}})^{\top}+O(\lambda),
\end{gather*}
as $|\lambda|\rightarrow0$, and
\begin{gather*}
\nabla\gamma^{(2)}(l)\sim(0,-1)^{\top}+(-u_{x_{2}x_{2}},u_{x_{1}x_{2}})^{\top}\lambda^{-1}+O\big(\lambda^{-2}\big),
\end{gather*}
as $|\lambda|\rightarrow\infty$. If we set
\begin{gather*}
 \nabla h^{(y)}\big(\tilde{l}\big):=\big(\lambda^{-1}\nabla\gamma^{(1)}\big(\tilde{l}\big)\big)\big|_{-}=\frac{u_{yx_{2}}}{\lambda}\frac{\partial}{\partial x_{1}}-\frac{u_{yx_{1}}}{\lambda}\frac{\partial}{\partial x_{2}},\\
\nabla h^{(t)}\big(\tilde{l}\big):=\big(\lambda\nabla\gamma^{(2)}\big(\tilde{l}\big)\big)\big|_{+}=-u_{x_{2}x_{2}}\frac{\partial}{\partial x_{1}}+(u_{x_{1}x_{2}}-\lambda)\frac{\partial}{\partial x_{2}},
\end{gather*}
the commutativity condition \eqref{C2} of the vector fields \eqref{C2a} leads to the modified Pleba\'{n}ski heavenly equation~\cite{DoFe}
\begin{gather}
u_{yt}-u_{yx_{1}}u_{x_{2}x_{2}}+u_{yx_{2}}u_{x_{1}x_{2}}=0, \label{H3a}
\end{gather}
with the Lax--Sato representation given by the first-order partial differential equations
\begin{gather*}
\frac{\partial\psi }{\partial y}-\frac{u_{yx_{2}}}{\lambda}\frac{\partial\psi }{\partial x_{1}}+\frac{u_{yx_{1}}}{\lambda}\frac{\partial\psi }{\partial x_{2}}=0,\\
\frac{\partial\psi }{\partial t}-u_{x_{2}x_{2}}\frac{\partial\psi }{\partial x_{1}}+(u_{x_{1}x_{2}}-\lambda)\frac{\partial \psi}{\partial x_{2}}=0
\end{gather*}
for functions $\psi \in C^{2}\big(\mathbb{R}^{2}\times\mathbb{T}^{2}\times\mathbb{C};\mathbb{C}\big)$.

The differential-geometric form of the seed element \eqref{H3} is also dimension invariant subject to additional spatial variables of the torus $\mathbb{T}^{2}\times\mathbb{C}$, $n>2$, which poses a natural question of finding the corresponding multi-dimensional generalizations of the modified Pleba\'{n}ski heavenly equation~\eqref{H3a}.

If a seed element $\tilde{l}\in\tilde{\mathcal{G}}^{\ast}=\widetilde{{\rm diff}}\big(\mathbb{T}^{2k}\times\mathbb{C}\big)^{\ast}$ is chosen in the form~\eqref{H3}, where $u\in C^{2}\big(\mathbb{R}^{2}\times\mathbb{T}^{2k};\mathbb{R}\big)$, we have the following asymptotic expansions for gradients of~$2k$ independent Casimir functionals $\gamma^{(j)}\in \mathrm{I}\big(\tilde{\mathcal{G}}^{\ast}\big)$, where $\tilde{\mathcal{G}}^{\ast}=\widetilde{{\rm diff}}\big(\mathbb{T}^{2k}\times\mathbb{C}\big)^{\ast}$, $j=1,\dots, 2k$:
\begin{gather*}
\nabla\gamma^{(1)}(l)\sim(-u_{yx_{2}},u_{yx_{1}},\underbrace{0,\ldots,0}_{2k-2})^{\top}+O(\lambda),\\
\nabla\gamma^{(3)}(l)\sim(0,0,-u_{yx_{4}},u_{yx_{3}},\underbrace{0,\ldots,0}_{2k-4})^{\top}+O(\lambda),\\
\cdots\cdots\cdots\cdots\cdots\cdots\cdots\cdots\cdots\cdots\cdots\cdots\cdots\cdots\cdots\cdots\\
\nabla\gamma^{(2k-1)}(l)\sim(\underbrace{0,\ldots,0}_{2k-2},-u_{yx_{2k}},u_{yx_{2k-1}})^{\top}+O(\lambda),
\end{gather*}
as $|\lambda|\rightarrow0$, and
\begin{gather*}
\nabla\gamma^{(2)}(l)\sim(0,-1,\underbrace{0,\ldots,0}_{2k-2})^{\top}+(-u_{x_{2}x_{2}},u_{x_{1}x_{2}},\underbrace{0,\ldots,0}_{2k-2})^{\top}\lambda^{-1}+O\big(\lambda^{-2}\big),\\
\nabla\gamma^{(4)}(l)\sim(0,0,-u_{x_{4}x_{2}},u_{x_{3}x_{2}},\underbrace{0,\ldots,0}_{2k-4})^{\top}\lambda^{-1}+O\big(\lambda^{-2}\big),\\
\cdots\cdots\cdots\cdots\cdots\cdots\cdots\cdots\cdots\cdots\cdots\cdots\cdots\cdots\cdots\cdots\cdots\\
\nabla\gamma^{(2k)}(l)\sim(\underbrace{0,\ldots,0}_{2k-2},-u_{x_{2k}x_{2}},u_{x_{2k-1}x_{2}})^{\top}\lambda^{-1}+O\big(\lambda^{-2}\big),
\end{gather*}
as $|\lambda|\rightarrow\infty$. In the case, when
\begin{gather*}
 \nabla h^{(y)}\big(\tilde{l}\big):=-\big(\lambda^{-1}\big(\nabla\gamma^{(1)}\big(\tilde{l}\big)+\dots+\nabla\gamma^{(2k-1)}\big(\tilde{l}\big)\big)\big)\big|_{-}\\
\hphantom{\nabla h^{(y)}\big(\tilde{l}\big)}{}=\sum_{m=1}^{k}\left( \frac{u_{yx_{2m}}}{\lambda}\frac{\partial
}{\partial x_{2m-1}}-\frac{u_{yx_{2m-1}}}{\lambda}\frac{\partial}{\partial x_{2m}}\right) ,\\
\nabla h^{(t)}\big(\tilde{l}\big):=\big(\lambda\big(\nabla\gamma^{(2)}\big(\tilde{l}\big)+\dots+\nabla\gamma^{(2k)}\big(\tilde{l}\big)\big)\big)\big|_{+}\\
\hphantom{\nabla h^{(t)}\big(\tilde{l}\big)}{}=-u_{x_{2}x_{2}}\frac{\partial}{\partial x_{1}}+(u_{x_{1}x_{2}}-\lambda)\frac{\partial}{\partial x_{2}}-\sum_{m=2}^{k}\left( u_{x_{2m}x_{2}}\frac{\partial}{\partial x_{2m-1}}-u_{x_{2m-1}x_{2}}\frac{\partial}{\partial x_{2m}}\right) ,
\end{gather*}
the commutativity condition \eqref{C2} of the vector fields \eqref{C2a} leads to the following multi-di\-men\-sio\-nal analogs of the modified Pleba\'{n}ski heavenly equation
\begin{gather*}
u_{yt}-\sum_{m=1}^{k}(u_{yx_{2m}}u_{x_{2}x_{2m-1}}-u_{yx_{2m-1}}u_{x_{2} x_{2m}})=0.
\end{gather*}

\subsection{Husain heavenly equation and its generalizations}

A seed element $\tilde{l}\in\tilde{\mathcal{G}}^{\ast}\simeq\widetilde{{\rm diff}}\big(\mathbb{T}^{2k}\times\mathbb{C}\big)^{\ast}$ in the form
\begin{gather}
\tilde{l}=\frac{{\rm d}(u_{y}+{\rm i}u_{t})}{\lambda-{\rm i}}+\frac{{\rm d}(u_{y}-{\rm i}u_{t})}{\lambda +{\rm i}}=\frac{2(\lambda {\rm d}u_{y}-{\rm d}u_{t})}{\lambda^{2}+1}, \label{H10}
\end{gather}
where ${\rm i}^{2}=-1$, $u\in C^{2}\big(\mathbb{R}^{2}\times\mathbb{T}^{2};\mathbb{R}\big)$, $(x_{1},x_{2})\in\mathbb{T}^{2}$, $\lambda\in\mathbb{C}\backslash\{-{\rm i};{\rm i}\}$, generates two independent Casimir functionals $\gamma^{(1)}$ and $\gamma^{(2)}\in \mathrm{I}\big(\tilde{\mathcal{G}}^{\ast}\big)$, with the following gradient asymptotic expansions
\begin{gather*}
\nabla\gamma^{(1)}(l)\sim\frac{1}{2}(-u_{yx_{2}}-{\rm i}u_{tx_{2}},u_{yx_{1}}+{\rm i}u_{tx_{1}})^{\top}+O(\mu),\qquad \mu:=\lambda-{\rm i},
\end{gather*}
as $|\mu|\rightarrow0$, and
\begin{gather*}
\nabla\gamma^{(2)}(l)\sim\frac{1}{2}(-u_{yx_{2}}+{\rm i}u_{tx_{2}},u_{yx_{1}}-{\rm i}u_{tx_{1}})^{\top}+O(\xi),\qquad \xi:=\lambda+{\rm i},
\end{gather*}
as $|\xi|\rightarrow0$. When
\begin{gather*}
 \nabla h^{(y)}\big(\tilde{l}\big):=\big(\mu^{-1}\nabla\gamma^{(1)}\big(\tilde{l}\big)+\xi^{-1}\nabla\gamma^{(2)}\big(\tilde{l}\big)\big)\big|_{-}\\
\hphantom{\nabla h^{(y)}\big(\tilde{l}\big)}{}=\frac{1}{2\mu}\left( (-u_{yx_{2}}-{\rm i}u_{tx_{2}})\frac{\partial}{\partial x_{1}}+(u_{yx_{1}}+{\rm i}u_{tx_{1}})\frac{\partial}{\partial x_{2}}\right) \\
\hphantom{\nabla h^{(y)}\big(\tilde{l}\big):=}{} +\frac{1}{2\xi}\left( (-u_{yx_{2}}+{\rm i}u_{tx_{2}})\frac{\partial}{\partial x_{1}}+(u_{yx_{1}}-{\rm i}u_{tx_{1}})\frac{\partial}{\partial x_{2}}\right) \\
\hphantom{\nabla h^{(y)}\big(\tilde{l}\big)}{}
 =\frac{u_{tx_{2}}-\lambda u_{yx_{2}}}{\lambda^{2}+1}\frac{\partial}{\partial x_{1}}+\frac{\lambda u_{yx_{1}}-u_{tx_{1}}}{\lambda^{2}+1}\frac{\partial}{\partial x_{2}},\\
 \nabla h^{(t)}\big(\tilde{l}\big):=\big({-}\mu^{-1}{\rm i}\nabla\gamma^{(1)}\big(\tilde{l}\big)+\xi^{-1}{\rm i}\nabla\gamma^{(2)}\big(\tilde{l}\big)\big)\big|_{-} \\
\hphantom{\nabla h^{(t)}\big(\tilde{l}\big)}{}
=\frac{1}{2\mu}\left( (-u_{tx_{2}}+{\rm i}u_{yx_{2}})\frac{\partial}{\partial x_{1}}+(u_{tx_{1}}-{\rm i}u_{yx_{1}})\frac{\partial}{\partial x_{2}}\right) \\
\hphantom{\nabla h^{(t)}\big(\tilde{l}\big):=}{}
 +\frac{1}{2\xi}\left( -(u_{tx_{2}}+{\rm i}u_{yx_{2}})\frac{\partial}{\partial x_{1}}+(u_{tx_{1}}+{\rm i}u_{yx_{1}})\frac{\partial}{\partial x_{2}}\right) \\
\hphantom{\nabla h^{(t)}\big(\tilde{l}\big)}{}
=-\frac{u_{yx_{2}}+\lambda u_{tx_{2}}}{\lambda^{2}+1}\frac{\partial}{\partial x_{1}}+\frac{u_{yx_{1}}+\lambda u_{tx_{1}}}{\lambda^{2}+1}\frac{\partial}{\partial x_{2}},
\end{gather*}
the commutativity condition \eqref{C2} of the vector fields \eqref{C2a} leads to the Husain heavenly equation~\cite{DoFe,ShYa}
\begin{gather}
u_{yy}+u_{tt}+u_{yx_{1}}u_{tx_{2}}-u_{yx_{2}}u_{tx_{1}}=0, \label{H11}
\end{gather}
with the Lax--Sato representation given by the first-order partial differential equations
\begin{gather*}
\frac{\partial\psi }{\partial y}+\frac{u_{tx_{2}}-\lambda u_{yx_{2}}}{\lambda^{2}+1}\frac{\partial\psi }{\partial x_{1}}+\frac{\lambda u_{yx_{1}}-u_{tx_{1}}}{\lambda^{2}+1}\frac{\partial \psi }{\partial x_{2}}=0,\\
\frac{\partial\psi }{\partial t}-\frac{u_{yx_{2}}+\lambda u_{tx_{2}}}{\lambda^{2}+1}\frac{\partial\psi }{\partial x_{1}}+\frac{u_{yx_{1}}+\lambda u_{tx_{1}}}{\lambda^{2}+1}\frac{\partial\psi }{\partial x_{2}}=0,
\end{gather*}
where $\psi \in C^{2}\big(\mathbb{R}^{2}\times\mathbb{T}^{2}\times\mathbb{C};\mathbb{C}\big)$.

The differential-geometric form of the seed element \eqref{H10} is also dimension invariant subject to additional spatial variables of the torus $\mathbb{T}^{n}$, $n>2$, which poses a natural question of finding the corresponding multi-dimensional generalizations of the Husain heavenly equation \eqref{H11}.

If a seed element $\tilde{l}\in\tilde{\mathcal{G}}^{\ast}=\widetilde{{\rm diff}}\big(\mathbb{T}^{2k}\times\mathbb{C}\big)^{\ast}$ is chosen in the form~\eqref{H10}, where $u\in C^{2}\big(\mathbb{R}^{2}\times\mathbb{T}^{2k};\mathbb{R}\big)$, we have the following asymptotic expansions for gradients of $2k$ independent Casimir functionals $\gamma^{(j)}\in \mathrm{I}\big(\tilde{\mathcal{G}}^{\ast}\big)$, where $\tilde{\mathcal{G}}^{\ast}=\widetilde{{\rm diff}}\big(\mathbb{T}^{2k}\times\mathbb{C}\big)^{\ast}$, $j= 1,\dots,2k$:
\begin{gather*}
\nabla\gamma^{(1)}(l)\sim\frac{1}{2}(-u_{yx_{2}}-{\rm i}u_{tx_{2}},u_{yx_{1} }+{\rm i}u_{tx_{1}},\underbrace{0,\ldots,0}_{2k-2})^{\top}+O(\mu),\\
 \nabla\gamma^{(3)}(l)\sim\frac{1}{2}(0,0,-u_{yx_{4}}-{\rm i}u_{tx_{4}},u_{yx_{3} }+{\rm i}u_{tx_{3}},\underbrace{0,\ldots,0}_{2k-4})^{\top}+O(\mu),\\
\cdots\cdots\cdots\cdots\cdots\cdots\cdots\cdots\cdots\cdots\cdots\cdots\cdots\cdots\cdots\cdots\cdots\cdots\cdots\cdots\cdots\cdots\\
 \nabla\gamma^{(2k-1)}(l)\sim\frac{1}{2}(\underbrace{0,\ldots,0}_{2k-2},-u_{yx_{2k}}-{\rm i}u_{tx_{2k}},u_{yx_{2k-1}}+{\rm i}u_{tx_{2k-1}})^{\top}+O(\mu),
\end{gather*}
as $|\mu|\rightarrow0$, and
\begin{gather*}
 \nabla\gamma^{(2)}(l)\sim\frac{1}{2}(-u_{yx_{2}}+{\rm i}u_{tx_{2}},u_{yx_{1}}-{\rm i}u_{tx_{1}},\underbrace{0,\ldots,0}_{2k-2})^{\top}+O(\xi),\\
\nabla\gamma^{(4)}(l)\sim\frac{1}{2}(0,0,-u_{yx_{4}}+{\rm i}u_{tx_{4}},u_{yx_{3}}-{\rm i}u_{tx_{3}},\underbrace{0,\ldots,0}_{2k-4})^{\top}+O(\xi),\\
\cdots\cdots\cdots\cdots\cdots\cdots\cdots\cdots\cdots\cdots\cdots\cdots\cdots\cdots\cdots\cdots\cdots\cdots\cdots\cdots\cdots \\
\nabla\gamma^{(2k)}(l)\sim\frac{1}{2}(\underbrace{0,\ldots,0}_{2k-2},-u_{yx_{2k}}+{\rm i}u_{tx_{2k}},u_{yx_{2k-1}}-{\rm i}u_{tx_{2k-1}})^{\top}+O(\xi),
\end{gather*}
as $|\xi|\rightarrow0$. For the case when
\begin{gather*}
 \nabla h^{(y)}\big(\tilde{l}\big):=\sum_{m=1}^{k}\big(\mu^{-1}\nabla\gamma^{(2m-1)}\big(\tilde{l}\big)+\xi^{-1}\nabla\gamma^{(2m)}\big(\tilde{l}\big)\big)\big|_{-}\\
\hphantom{\nabla h^{(y)}\big(\tilde{l}\big)}{} =\sum_{m=1}^{k}\left( \frac{u_{tx_{2m}}-\lambda u_{yx_{2m}}
}{\lambda^{2}+1}\frac{\partial}{\partial x_{2m-1}}+\frac{\lambda u_{yx_{2m-1}
}-u_{tx_{2m-1}}}{\lambda^{2}+1}\frac{\partial}{\partial x_{2m}}\right) ,\\
 \nabla h^{(t)}\big(\tilde{l}\big):=\sum_{m=1}^{k}{\rm i}\big({-}\mu^{-1}\nabla\gamma^{(2m-1)}\big(\tilde{l}\big)+\xi^{-1}\nabla\gamma^{(2m)}\big(\tilde{l}\big)\big)\big|_{-}\\
\hphantom{\nabla h^{(t)}\big(\tilde{l}\big)}{} =\sum_{m=1}^{k}\left( -\frac{u_{yx_{2m}}+\lambda u_{tx_{2m}}}{\lambda^{2}+1}\frac{\partial}{\partial x_{2m-1}}+\frac{u_{yx_{2m-1}}+\lambda u_{tx_{2m-1}}}{\lambda^{2}+1}\frac{\partial}{\partial x_{2m}}\right) ,
\end{gather*}
the commutativity condition \eqref{C2} of the vector fields \eqref{C2a} leads to the following multi-di\-men\-sio\-nal analogs of the Husain heavenly equation
\begin{gather*}
u_{yy}+u_{tt}+\sum_{m=1}^{k}(u_{yx_{2m-1}}u_{tx_{2m}}-u_{yx_{2m}}u_{x_{2}x_{2m-1}})=0.
\end{gather*}

\subsection{The general Monge heavenly equation and its generalizations}

The seed element $\tilde{l}\in\tilde{\mathcal{G}}^{\ast}=\widetilde{{\rm diff}}\big(\mathbb{T}^{4}\times\mathbb{C}\big)^{\ast}$, taken in the form
\begin{gather*}
\tilde{l}={\rm d}u_{y}+\lambda^{-1}({\rm d}x_{1}+{\rm d}x_{2}),
\end{gather*}
where $u\in C^{2}\big(\mathbb{R}^{2}\times\mathbb{T}^{4};\mathbb{R}\big)$, $(x_{1},x_{2},x_{3},x_{4})\in\mathbb{T}^{4}$, $\lambda\in\mathbb{C}\backslash\{0\}$, generates four independent Casimir functionals $\gamma^{(1)}$, $\gamma^{(2)}$, $\gamma^{(3)}$ and $\gamma^{(4)}\in\mathrm{I}\big(\tilde{\mathcal{G}}^{\ast}\big)$, with gradients having the asymptotic expansions
\begin{gather*}
 \nabla\gamma^{(1)}(l)\sim(0,1,0,0)^{\top}\\
 \hphantom{\nabla\gamma^{(1)}(l)\sim}{} +\big({-}u_{yx_{2}}-(\partial_{x_{2}}-\partial_{x_{1}})^{-1}u_{yx_{2}x_{1}},(\partial_{x_{2}} -\partial_{x_{1}})^{-1}u_{yx_{2}x_{1}},0,0\big)^{\top}\lambda+O\big(\lambda^{2}\big),\\
 \nabla\gamma^{(2)}(l)\sim(1,0,0,0)^{\top}\\
 \hphantom{\nabla\gamma^{(2)}(l)\sim}{} + \big((\partial_{x_{1}}-\partial_{x_{2}})^{-1}u_{yx_{1}x_{2}},-u_{yx_{1}}-(\partial_{x_{1}}-\partial_{x_{2}})^{-1}u_{yx_{1}x_{2}
},0,0\big)^{\top}\lambda+O\big(\lambda^{2}\big),\\
 \nabla\gamma^{(3)}(l)\sim(0,0,-u_{yx_{4}},u_{yx_{3}})^{\top}+O\big(\lambda^{2}\big),\\
 \nabla\gamma^{(4)}(l)\sim(0,0,-u_{tx_{4}},u_{tx_{3}})^{\top}\\
 \hphantom{\nabla\gamma^{(4)}(l)\sim}{} +(u_{yx_{3}}u_{tx_{4}}-u_{yx_{4}}u_{tx_{3}},0, u_{yx_{4}}u_{tx_{1}}-u_{yx_{1}}u_{tx_{4}},u_{yx_{1}}u_{tx_{3}
}-u_{yx_{3}}u_{tx_{1}})^{\top}\lambda+O\big(\lambda^{2}\big),
\end{gather*}
as $|\lambda|\rightarrow0$. When
\begin{gather*}
 \nabla h^{(y)}\big(\tilde{l}\big):=\big(\lambda^{-1}\big(\nabla\gamma^{(1)}\big(\tilde{l}\big)+\nabla\gamma^{(3)}\big(\tilde{l}\big)\big)\big)\big|_{-}
 =0\frac{\partial}{\partial x_{1}}+\frac{1}{\lambda}\frac{\partial
}{\partial x_{2}}-\frac{u_{yx_{4}}}{\lambda}\frac{\partial}{\partial x_{3}}+\frac{u_{yx_{3}}}{\lambda}\frac{\partial}{\partial x_{4}},\\
 \nabla h^{(t)}\big(\tilde{l}\big):=\big(\lambda^{-1}\big({-}\nabla\gamma^{(2)}\big(\tilde{l}\big)+\nabla\gamma^{(4)}\big(\tilde{l}\big)\big)\big)\big|_{-}
 =-\frac{1}{\lambda}\frac{\partial}{\partial x_{1}}+0\frac{\partial}{\partial x_{2}}-\frac{u_{tx_{4}}}{\lambda}\frac{\partial}{\partial x_{3}}+\frac{u_{tx_{3}}}{\lambda}\frac{\partial}{\partial x_{4}},
\end{gather*}
the commutativity condition \eqref{C2} of the vector fields \eqref{C2a} leads to the general Monge heavenly equation~\cite{DoFeKrNo}
\begin{gather*}
u_{yx_{1}}+u_{tx_{2}}+u_{yx_{3}}u_{tx_{4}}-u_{yx_{4}}u_{tx_{3}}=0,
\end{gather*}
with the Lax--Sato representation given by the first-order partial differential equations
\begin{gather*}
 \frac{\partial\psi}{\partial y}+\frac{1}{\lambda}\frac{\partial\psi }{\partial x_{2}}-\frac{u_{yx_{4}}}{\lambda}\frac{\partial\psi}{\partial x_{3}}+\frac{u_{yx_{3}}}{\lambda}\frac{\partial\psi}{\partial x_{4}}=0,\\
 \frac{\partial\psi}{\partial t}-\frac{1}{\lambda}\frac{\partial\psi}{\partial x_{1}}-\frac{u_{tx_{4}}}{\lambda}\frac{\partial\psi}{\partial x_{3}}+\frac{u_{tx_{3}}}{\lambda}\frac{\partial\psi}{\partial x_{4}}=0,
\end{gather*}
where $\psi\in C^{2}\big(\mathbb{R}^{2}\times\mathbb{T}^{4}\times\mathbb{C};\mathbb{C}\big)$ and $\lambda\in\mathbb{C}\backslash\{0\}$.

Taking into account that the condition for Casimir invariants is equivalent to a system of homogeneous linear first-order partial differential equations for a covector function $l=(l_{1},l_{2},l_{3},l_{4})^{\top}$, the corresponding seed element can be chosen in different forms. For example, if the expression
\begin{gather*}
\tilde{l}={\rm d}u_{t}+\lambda^{-1}({\rm d}x_{1}+{\rm d}x_{2})
\end{gather*}
is considered as a seed element, it generates four independent Casimir functionals $\gamma^{(1)}$, $\gamma^{(2)}$, $\gamma^{(3)}$ and $\gamma ^{(4)}\in \mathrm{I}\big(\tilde{\mathcal{G}}^{\ast}\big)$, whose gradients have the following asymptotic expansions
\begin{gather*}
 \nabla\gamma^{(1)}(l)\sim(0,1,0,0)^{\top}\\
 \hphantom{\nabla\gamma^{(1)}(l)\sim}{} +
\big({-}u_{tx_{2}}-(\partial_{x_{2}}-\partial_{x_{1}})^{-1}u_{tx_{2}x_{1}},(\partial_{x_{2}}-\partial_{x_{1}})^{-1}u_{tx_{2}x_{1}},0,0\big)^{\top}\lambda+O\big(\lambda^{2}\big),\\
 \nabla\gamma^{(2)}(l)\sim(1,0,0,0)^{\top}\\
 \hphantom{\nabla\gamma^{(2)}(l)\sim}{} +\big((\partial_{x_{1}}-\partial_{x_{2}})^{-1}u_{tx_{1}x_{2}},-u_{tx_{1}}-(\partial_{x_{1}} -\partial_{x_{2}})^{-1}u_{tx_{1}x_{2}},0,0\big)^{\top}\lambda+O\big(\lambda^{2}\big),\\
 \nabla\gamma^{(3)}(l)\sim(0,0,-u_{tx_{4}},u_{tx_{3}})^{\top}\\
 \hphantom{\nabla\gamma^{(3)}(l)\sim}{} +(0,u_{tx_{3}}u_{yx_{4}}-u_{tx_{4}}u_{yx_{3}},
 u_{tx_{4}}u_{yx_{2}}-u_{tx_{2}}u_{yx_{4}},u_{tx_{2}}u_{yx_{3}}-u_{tx_{3}}u_{yx_{2}})^{\top}\lambda+O\big(\lambda^{2}\big),\\
 \nabla\gamma^{(4)}(l)\sim(0,0,-u_{yx_{4}},u_{yx_{3}})^{\top}+O\big(\lambda^{2}\big),
\end{gather*}
as $|\lambda|\rightarrow0$. If a seed element has the form
\begin{gather}
\tilde{l}={\rm d}u_{y}+{\rm d}u_{t}+\lambda^{-1}({\rm d}x_{1}+{\rm d}x_{2}), \label{Hen**}%
\end{gather}
the asymptotic expansions for gradients of four independent Casimir functionals $\gamma^{(1)},\gamma^{(2)},\gamma^{(3)}$ and $\gamma^{(4)}\in \mathrm{I}\big(\tilde{\mathcal{G}}^{\ast}\big)$ are written as
\begin{gather*}
 \nabla\gamma^{(1)}(l)\sim(0,1,0,0)^{\top}+\big({-}(u_{yx_{2}}+u_{tx_{2}})-(\partial_{x_{2}}-\partial_{x_{1}})^{-1}(u_{yx_{2}x_{1}}+u_{tx_{2}x_{1}}),\\
\hphantom{\nabla\gamma^{(1)}(l)\sim}{} (\partial_{x_{2}}-\partial_{x_{1}})^{-1}(u_{yx_{2}x_{1}}+u_{tx_{2}x_{1}}),0,0\big)^{\top}\lambda+O\big(\lambda^{2}\big),\\
 \nabla\gamma^{(2)}(l)\sim(1,0,0,0)^{\top}+\big((\partial_{x_{1}}-\partial_{x_{2}})^{-1}(u_{yx_{1}x_{2}}+u_{tx_{1}x_{2}}),\\
\hphantom{\nabla\gamma^{(2)}(l)\sim}{} {-}(u_{yx_{1}}+u_{tx_{1}})-(\partial_{x_{1}}-\partial_{x_{2}})^{-1}(u_{yx_{1}x_{2}}+u_{tx_{1}x_{2}}),0,0\big)^{\top}\lambda+O\big(\lambda^{2}\big),\\
\nabla\gamma^{(3)}(l)\sim(0,0,-u_{yx_{4}},u_{yx_{3}})^{\top}+\big(0,u_{tx_{3}}u_{yx_{4}}-u_{tx_{4}}u_{yx_{3}},\\
\hphantom{\nabla\gamma^{(3)}(l)\sim}{} u_{tx_{4}}u_{yx_{2}}-u_{tx_{2}}u_{yx_{4}},u_{tx_{2}}u_{yx_{3}}-u_{tx_{3}}u_{yx_{2}}\big)^{\top}\lambda+O\big(\lambda^{2}\big),\\
\nabla\gamma^{(4)}(l)\sim(0,0,-u_{tx_{4}},u_{tx_{3}})^{\top}+\big(u_{yx_{3}}u_{tx_{4}}-u_{yx_{4}}u_{tx_{3}},0,\\
\hphantom{\nabla\gamma^{(4)}(l)\sim}{} u_{yx_{4}}u_{tx_{1}}-u_{yx_{1}}u_{tx_{4}},u_{yx_{1}}u_{tx_{3}}-u_{yx_{3}}u_{tx_{1}}\big)^{\top}\lambda+O\big(\lambda^{2}\big),
\end{gather*}
as $|\lambda|\rightarrow0$.

The above scheme can be generalized for all $n=2k$, where $k\in\mathbb{N}$, and $n>2$. In this case one has $2k$ independent Casimir functionals $\gamma^{(j)}\in\mathrm{I}\big(\tilde{\mathcal{G}}^{\ast}\big)$, where $\tilde{\mathcal{G}}^{\ast}=\widetilde{{\rm diff}}\big(\mathbb{T}^{2k}\times \mathbb{C}\big)^{\ast}$, $j= 1,\dots,2k$, whose gradient asymptotic expansions are equal to the following expressions
\begin{gather*}
 \nabla\gamma^{(1)}(l)\sim(0,1,\underbrace{0,\ldots,0}_{2k-2})^{\top
}+\big({-}(u_{yx_{2}}+u_{tx_{2}})-(\partial_{x_{2}}-\partial_{x_{1}})^{-1}%
(u_{yx_{2}x_{1}}+u_{tx_{2}x_{1}}),\\
\hphantom{\nabla\gamma^{(1)}(l)\sim} {}(\partial_{x_{2}}-\partial_{x_{1}})^{-1}(u_{yx_{2}x_{1}}+u_{tx_{2}x_{1}}),\underbrace{0,\ldots,0}_{2k-2}\big)^{\top}\lambda+O\big(\lambda^{2}\big),\\
 \nabla\gamma^{(2)}(l)\sim(1,0,\underbrace{0,\ldots,0}_{2k-2})^{\top
}+\big((\partial_{x_{1}}-\partial_{x_{2}})^{-1}(u_{yx_{1}x_{2}}+u_{tx_{1}x_{2}}),\\
 \hphantom{\nabla\gamma^{(2)}(l)\sim}{}{-}(u_{yx_{1}}+u_{tx_{1}})-(\partial_{x_{1}}-\partial_{x_{2}}%
)^{-1}(u_{yx_{1}x_{2}}+u_{tx_{1}x_{2}}),\underbrace{0,\ldots,0}_{2k-2}\big)^{\top
}\lambda+O\big(\lambda^{2}\big),\\
 \nabla\gamma^{(3)}(l)\sim(0,0,-u_{yx_{4}},u_{yx_{3}},\underbrace{0,\ldots,0}_{2k-4})^{\top}+(0,u_{tx_{3}}u_{yx_{4}}-u_{tx_{4}}u_{yx_{3}},\\
\hphantom{\nabla\gamma^{(3)}(l)\sim} {}u_{tx_{4}}u_{yx_{2}}-u_{tx_{2}}u_{yx_{4}},u_{tx_{2}}u_{yx_{3}}-u_{tx_{3}}u_{yx_{2}},\underbrace{0,\ldots,0}_{2k-4})^{\top}\lambda
+O\big(\lambda^{2}\big),\\
 \nabla\gamma^{(4)}(l)\sim(0,0,-u_{tx_{4}},u_{tx_{3}},\underbrace{0,\ldots,0}_{2k-4})^{\top}+(u_{yx_{3}}u_{tx_{4}}-u_{yx_{4}}u_{tx_{3}},0,\\
\hphantom{\nabla\gamma^{(4)}(l)\sim} {}u_{yx_{4}}u_{tx_{1}}-u_{yx_{1}}u_{tx_{4}},u_{yx_{1}}u_{tx_{3}
}-u_{yx_{3}}u_{tx_{1}},\underbrace{0,\ldots,0}_{2k-4})^{\top}\lambda+O\big(\lambda^{2}\big),\\
\cdots\cdots\cdots\cdots\cdots\cdots\cdots\cdots\cdots\cdots\cdots\cdots\cdots\cdots\cdots\cdots\cdots\cdots \cdots\cdots\cdots\cdots\cdots\cdots\\
 \nabla\gamma^{(2k-1)}(l)\sim(\underbrace{0,\ldots,0}_{2k-4},0,0,-u_{yx_{2k}},u_{yx_{2k-1}})^{\top} +(\underbrace{0,\ldots,0}_{2k-4},0,u_{tx_{2k-1}}u_{yx_{2k}}-u_{tx_{2k}}u_{yx_{2k-1}},\\
\hphantom{\nabla\gamma^{(2k-1)}(l)\sim} {}u_{tx_{2k}}u_{yx_{2}}-u_{tx_{2}}u_{yx_{2k}},u_{tx_{2}}u_{yx_{2k-1}}-u_{tx_{2k-1}}u_{yx_{2}})^{\top}\lambda+O\big(\lambda^{2}\big),\\
 \nabla\gamma^{(2k)}(l)\sim(\underbrace{0,\ldots,0}_{2k-4},0,0,-u_{tx_{2k}},u_{tx_{2k-1}})^{\top} +(\underbrace{0,\ldots,0}_{2k-4},u_{yx_{2k-1}}u_{tx_{2k}}-u_{yx_{2k}}u_{tx_{2k-1}},0,\\
\hphantom{\nabla\gamma^{(2k)}(l)\sim} {}u_{yx_{2k}}u_{tx_{1}}-u_{yx_{1}}u_{tx_{2k}},u_{yx_{1}}u_{tx_{2k-1}}-u_{yx_{2k-1}}u_{tx_{1}})^{\top}\lambda+O\big(\lambda^{2}\big),
\end{gather*}
when a seed element $\tilde{l}\in\tilde{\mathcal{G}}^{\ast}$ is chosen as in~\eqref{Hen**}. If
\begin{gather*}
 \nabla h^{(y)}\big(\tilde{l}\big):=\big(\lambda^{-1}\big(\nabla\gamma^{(1)}\big(\tilde{l}\big)+\nabla\gamma^{(3)}\big(\tilde{l}\big)+\dots +\nabla\gamma^{(2k-1)}\big(\tilde{l}\big)\big)\big)\big|_{-} \\
\hphantom{\nabla h^{(y)}\big(\tilde{l}\big)}{}=0\frac{\partial}{\partial x_{1}}+\frac{1}{\lambda}\frac{\partial}{\partial x_{2}}-\sum_{m=2}^{k}\left( \frac{u_{yx_{2m}}}{\lambda}%
\frac{\partial}{\partial x_{2m-1}}-\frac{u_{yx_{2m-1}}}{\lambda}\frac{\partial}{\partial x_{2m}}\right) ,\\
 \nabla h^{(t)}\big(\tilde{l}\big):=\big(\lambda^{-1}\big({-}\nabla\gamma^{(2)}\big(\tilde{l}\big)+\nabla\gamma^{(4)}\big(\tilde{l}\big)+\dots +\nabla\gamma^{(2k)}\big(\tilde{l}\big)\big)\big)\big|_{-} \\
\hphantom{\nabla h^{(t)}\big(\tilde{l}\big)}{}=-\frac{1}{\lambda}\frac{\partial}{\partial x_{1}}+0\frac{\partial}{\partial x_{2}}-\sum_{m=2}^{k}\left( \frac{u_{tx_{2m}}}{\lambda}
\frac{\partial}{\partial x_{2m-1}}-\frac{u_{tx_{2m-1}}}{\lambda}\frac{\partial}{\partial x_{2m}}\right) ,
\end{gather*}
the commutativity condition \eqref{C2} of the vector fields \eqref{C2a} leads to the following multi-di\-men\-sio\-nal analogs of the general Monge heavenly equation
\begin{gather*}
u_{yx_{1}}+u_{tx_{2}}+\sum_{m=2}^{k}(u_{yx_{2m-1}}u_{tx_{2m}}-u_{yx_{2m}}u_{tx_{2m-1}})=0.
\end{gather*}

\subsection{Superanalogs of the Whitham heavenly equation}

Assume now that an element $\tilde{l}\in\tilde{\mathcal{G}}^{\ast}$, where $\widetilde{\mathcal{G}}:=\widetilde{{\rm diff}}\big(\mathbb{T}^{1|N}\times \mathbb{C}\big)=\widetilde{{\rm diff}}_{+}\big(\mathbb{T}^{1|N}\times\mathbb{C}\big)\oplus\widetilde{{\rm diff}}\big(\mathbb{T}^{1|N}\times\mathbb{C}\big)$ is the loop Lie algebra of the superconformal diffeomorphism group $\widetilde{{\rm Diff}}\big(\mathbb{T}^{1|N}\times\mathbb{C}\big)$ of vector fields on a super-$1|N$-dimensional set $\mathbb{T}^{1|N}\times\mathbb{C}:=\mathbb{T}^{1}\times \Lambda_{1}^{N}\times\mathbb{C}$ (see~\cite{HenPryk-super}), constructed over a finite-dimensional Grassmann algebra $\Lambda:=\Lambda_{0}\oplus\Lambda_{1}$ over $\mathbb{C}$, $\Lambda_{0}\supset\mathbb{C}$, admits the following asymptotic expansions for gradients of the Casimir invariants $h^{(1)},h^{(2)}\in \mathrm{I}\big(\tilde{\mathcal{G}}^{\ast}\big)$:
\begin{gather}
\nabla h^{(1)}(l)\sim w_{y}+O(\lambda) \label{Hen_sec7_1}
\end{gather}
as $|\lambda|\rightarrow0$, and
\begin{gather}
\nabla h^{(2)}(l)\sim 1-w_{x}\lambda^{-1}+O\big(\lambda^{-2}\big) \label{Hen_sec7_2}
\end{gather}
as $|\lambda|\rightarrow\infty$. Then the commutativity condition for the Hamiltonian flows
\begin{alignat*}{3}
& {\rm d}\tilde{l}/{\rm d}y=\operatorname{ad}_{\nabla h_{-}^{(y)}(\tilde{l})}^{\ast}\tilde{l},\qquad && \nabla h_{-}^{(y)}(l)=-\big(\lambda^{-1}\nabla h^{1}(l)\big)_{-}
=-w_{y}\lambda^{-1},& \\
& {\rm d}\tilde{l}/{\rm d}t=-\operatorname{ad}_{\nabla h_{+}^{(t)}(\tilde{l})}^{\ast}\tilde{l},\qquad && \nabla h_{+}^{(t)}(l)=-\big(\lambda\nabla h^{(2)}(l)\big)_{+}
=-\lambda+w_{x},&
\end{alignat*}
naturally leads to the heavenly type equation
\begin{gather}
w_{yt}=w_{x}w_{yx}-w_{y}w_{xx}-\frac{1}{2}\sum_{i=1}^{N}(D_{\vartheta_{i}}w_{x})(D_{\vartheta_{i}}w_{y}), \label{Hen_sec8c_1}%
\end{gather}
where $w\in C^{2}\big(\mathbb{R}^{2}\times\mathbb{T}^{1|N}\times\mathbb{C};\Lambda_{0}\big)$ and $D_{\vartheta_{i}}:=\partial/\partial\vartheta _{i}+\vartheta_{i}\partial/\partial x$, $i= 1,\dots,N$, are superderivatives with respect to the anticommuting variables $\vartheta_{i}\in\Lambda_{1}$, $i=1,\dots,N$.

This equation can be considered as a super-generalization of the Whitham heavenly one \cite{HePrBlPr,MaMe,Moroz1} for every~$N\in\mathbb{N}$. The compatibility condition for the first-order partial differential equations
\begin{gather*}
 \psi_{y}+\frac{1}{\lambda}\left( w_{y}\psi_{x}+\frac{1}{2}\sum_{i=1}^{N}(D_{\vartheta_{i}}w_{y})(D_{\vartheta_{i}}\psi)\right) =0,\\
 \psi_{t}+(-\lambda+w_{x})\psi_{x}+\frac{1}{2}\sum_{i=1}^{N}(D_{\vartheta _{i}}w_{x})(D_{\vartheta_{i}}\psi)=0,
\end{gather*}
where $\psi\in C^{2}\big(\mathbb{R}^{2}\times\mathbb{T}^{1|N}\times\mathbb{C};\Lambda_{0}\big)$ and $\lambda\in\mathbb{C}\backslash\{0\}$, give rise to the corresponding Lax--Sato representation of the heavenly type equation~\eqref{Hen_sec8c_1}.

Moreover, based on straightforward calculations, one can obtain from the Casimir invariant equation the corresponding seed element $\tilde{l}:=l{\rm d}x\in\tilde{\mathcal{G}}^{\ast}$, which can be written in the following form for an arbitrary $N\in\mathbb{N}$:
\begin{gather*}
l=Ca^{-\frac{4-N}{2}},\qquad a:=\nabla h(l),
\end{gather*}
where a scalar function $C=C(x;\vartheta)$ satisfies a linear homogeneous ordinary differential equation
\begin{gather*}
C_{x}=\langle DC,Q\rangle,
\end{gather*}
$Q=(Q_{1},\ldots,Q_{N})$, $Q_{i}=\frac{(-1)^{N}}{2}(D_{\vartheta_{i}}\ln a)$, in the superspace ${\mathbb{R}}^{2^{N-1}|2^{N-1}} \simeq\Lambda_{0}^{2^{N-1}}\times\Lambda_{1}^{2^{N-1}}$. More\-over, $C\in C^{\infty}(\mathbb{T}^{1|N}\times\mathbb{C};\Lambda_{1})$ if $N$ is an odd natural number and $C\in C^{\infty}\big(\mathbb{T}_{\mathbb{C}}^{1|N};\Lambda_{0}\big)$, if $N$ is even. In particular, when $N=1$ one has
\begin{gather*}
l=C_{1}\big(\partial_{x}^{-1}D_{\theta_{1}}a^{-\frac{1}{2}}\big)a^{-\frac{3}{2}},
\end{gather*}
where $C_{1}\in\mathbb{R}$ is constant.

If $N=1$ and $C_{1}=1$, the corresponding seed-element $\tilde{l}\in \tilde{\mathcal{G}}^{\ast}$, related to the asymptotic expansions~\eqref{Hen_sec7_1} and~\eqref{Hen_sec7_2}, can be reduced to
\begin{gather*}
\tilde{l}=\big[\lambda^{-1}\big(\partial_{x}^{-1}D_{\theta_{1}}w_{y}^{-\frac{1}{2}}\big)w_{y}^{-\frac{3}{2}}+\xi_{x}/2+\theta_{1}(2u_{x}+\lambda)\big]{\rm d}x,
\end{gather*}
where $w:=u+\theta_{1}\xi$, $u\in C^{\infty}\big(\mathbb{R}^{2}\times \mathbb{T}^{1|1}\times\mathbb{C};\Lambda_{0}\big)$ and $\xi\in C^{\infty}\big(\mathbb{R}^{2}\times\mathbb{T}^{1|1}\times\mathbb{C};\Lambda_{1}\big)$.

\section{Conclusions}

We applied a Lie-algebraic approach (involving holomorphic extension) to studying vector fields on $\mathbb{T}^{n}\times\mathbb{C}$ and the related structures, which made it possible to construct a wide class of multi-dimensional dispersionless integrable systems describing conformal structure generating equations of modern mathematical physics. Also described was a modification of the approach subject to the spatial-dimensional invariance and meromorphicity of the related differential-geometric structures, giving rise to new generalized multi-dimensional conformal metric equations. We presented a detailed analysis of the related differential-geometric structures of the Einstein--Weyl conformal metric
equation, the modified Einstein--Weyl metric equation, the Dunajski heavenly equation system, the first and second conformal structure generating equations and the inverse first Shabat reduction heavenly equation. In addition, we analyzed the first and modified Pleba\'{n}ski heavenly equations, the Husain heavenly equation, the general Monge equation and their multi-dimensional generalizations as well as superconformal analogs of the Whitham heavenly equation.

\subsection*{Acknowledgements}

The authors are cordially indebted to Professors Alexander Balinsky, Maxim Pavlov and Artur Sergyeyev for useful comments and remarks, especially for elucidating references that were very instrumental in preparing this manuscript. They also are indebted to Professor Anatol Odzije\-wicz for fruitful and instructive discussions during the XXXVII Workshop on Geometric Methods in Physics held on July 1--7, 2018 in Bia{\l}owie\.{z}a, Poland. Thanks are also due the Department of Physics, Mathematics and Computer Science of the Cracow University of Technology for a~local research grant F-2/370/2018/DS. This work was partly funded by the budget program of Ukraine ``Support for the development of priority research areas'' (CPCEC 6451230). Last but not least, thanks are due to the referees for carefully reading our work, making insightful remarks and posing questions, which were useful when preparing a revised manuscript.

\pdfbookmark[1]{References}{ref}
\LastPageEnding

\end{document}